\documentclass[twocolumn,floatfix,showpacs,floats,aps,prb,showpacs,amsfonts,amssymb]{revtex4}
\usepackage{amssymb}
\usepackage{dcolumn}
\usepackage{bm}
\usepackage{hyphenat}
\usepackage{graphicx}
\usepackage{psfrag}
\usepackage[latin1]{inputenc}
\usepackage{hyperref}

\newlength{\sepmod}
\def\eqref#1{(\ref{#1})}
\def\l{\langle}

\def\r{\rangle}

\begin{document}
\title{Zero-temperature phase of the {\em XY\/} spin glass in two dimensions: Genetic
  embedded matching heuristic}

\author{Martin Weigel}
\altaffiliation{Present address: Institut f\"ur Physik,
  Johannes-Gutenberg-Universit\"at Mainz, Staudinger Weg 7, 55099 Mainz, Germany}
\email{weigel@uni-mainz.de}
\affiliation{Department of Mathematics and the Maxwell Institute for Mathematical
  Sciences, Heriot-Watt University, Edinburgh, EH14~4AS, UK}

\author{Michel J.\ P.\ Gingras}
\email{gingras@gandalf.uwaterloo.ca}
\affiliation{Department of Physics and Astronomy, University of Waterloo, Waterloo, Ontario,
  N2L~3G1, Canada}

\date{\today}

\begin{abstract} 

  For many real spin-glass materials, the Edwards-Anderson model with
  continuous-symmetry spins is more realistic than the rather better understood Ising
  variant.  In principle, the nature of an occurring spin-glass phase in such systems
  might be inferred from an analysis of the zero-temperature properties.
  Unfortunately, with few exceptions, the problem of finding ground-state
  configurations is a non-polynomial problem computationally, such that efficient
  approximation algorithms are called for. Here, we employ the recently developed
  genetic embedded matching (GEM) heuristic to investigate the nature of the
  zero-temperature phase of the bimodal {\em XY\/} spin glass in two dimensions. We
  analyze bulk properties such as the asymptotic ground-state energy and the phase
  diagram of disorder strength vs.\ disorder concentration. For the case of a
  symmetric distribution of ferromagnetic and antiferromagnetic bonds, we find that
  the ground state of the model is unique up to a {\em global} O($2$) rotation of the
  spins. In particular, there are no extensive degeneracies in this model. The main
  focus of this work is on an investigation of the excitation spectrum as probed by
  changing the boundary conditions. Using appropriate finite-size scaling techniques,
  we consistently determine the stiffness of spin and chiral domain walls and the
  corresponding fractal dimensions. Most noteworthy, we find that the spin and chiral
  channels are characterized by two distinct stiffness exponents and, consequently,
  the system displays spin-chirality decoupling at large length scales. Results for
  the overlap distribution do not support the possibility of a multitude of
  thermodynamic pure states.
\end{abstract}


\maketitle

\section{Introduction}

The discovery of materials without magnetic ordering down to zero temperature, but
with an unusual cusp in the non-linear magnetic susceptibility \cite{cannella:72}
indicating a transition to a magnetically frozen but random state, has prompted a
theoretical effort spanning over thirty years and aimed at the understanding of the
perplexing properties of these spin-glass systems
\cite{binder:86a,fischer:book,young:book,kawashima:03a}. Albeit mostly driven by the
theoreticians' hope to understand the peculiar behavior of such systems within the
framework of statistical mechanics, the results of this research have found
widespread application in seemingly rather distant fields such as the theory of
associative memory \cite{nishimori:book}, models of the immune system
\cite{parisi:90} or error correcting codes \cite{sourlas:89,dennis:02a}. Spin-glass
science thus provides for a classical example of how research driven by curiosity
rather than application can lead to unexpected progress in diverse research areas.
The canonical case of a spin glass, i.e., a noble metal host doped with magnetic
transition metal ions \cite{binder:86a} is not of particular interest as a material
apart from the peculiarity of its properties. In recent years, however, many new
materials with possible spin-glass phases have been discovered in compounds with
highly frustrating lattice structures \cite{greedan:01}.  The large degeneracy of the
ground state arising from geometric frustration alone gives rise to a host of novel
and exotic magnetic phases, including spin glasses \cite{gingras:96a}, but also spin
liquids \cite{gardner:99,mendels:07} and spin ices \cite{bramwell:01}. Experimental
evidence suggests that spin-glass like phenomena appear to lie at the heart of
several (potentially) technologically important phenomena, such as high temperature
superconductivity, colossal magnetoresistance, and the anomalous Hall effect.

Other than might be suspected from the pure volume of published research in the field
\cite{binder:86a,kawashima:03a}, however, owed to the unwieldy nature of the problem,
even some very basic questions such as the existence of a finite-temperature phase
transition have in general remained unanswered to date \cite{kawashima:03a}. Starting
from the, by now, rather complete \cite{talagrand:book} understanding of the
unusually complex mean-field theory \cite{parisi:79,parisi:80,parisi:80a} of the
fully connected (infinite dimensional) Sherrington-Kirkpatrick \cite{sherrington:75}
(SK) model, recent research has focused on the case of finite-dimensional systems.
For the nearest-neighbor O($n$) Edwards-Anderson (EA) model \cite{edwards:75a} with
Hamiltonian
\begin{equation}
  \label{eq:EA_model}
  {\cal H} = -\sum_{\l ij\r}J_{ij}\,\bm{S}_i\cdot\bm{S}_j,
\end{equation}
which has been studied in most detail, most of the effort has been devoted to the
discrete case of classical Ising spins ($n=1$). Even for this simplest case some
controversial questions, such as the existence of a glassy phase at
finite-temperatures in three dimensions (3D), could only be definitely answered in
very recent years \cite{kawashima:03a}. In general, many features such as the generic
structure of valleys and barriers in the free energy and, more generally, the
question of whether spin glasses in finite dimensions behave essentially alike or
differently from the mean-field case remain poorly understood
\cite{newman:03a,contucci:06}. Even the allegedly simplest case of the Ising spin
glass in two dimensions (2D) remains a topic of lively research activity
\cite{katzgraber:04,lukic:04,katzgraber:05a,jorg:06,fisch:06b}.

For most of the materials featuring glassy magnetic phases, however, continuous {\em
  XY\/} or Heisenberg spins certainly are a more accurate description than the Ising
model of extreme anisotropy in spin space
\cite{binder:86a,fischer:book,gingras:96a}. From the perspective of condensed matter
physics, much attention has been paid to the experimentally important case of weak,
but frustrating impurities and the question of a resulting loss of collinear
order. Isolated frustrating antiferromagnetic defects in an otherwise ferromagnetic
system couple to each other via an effective dipolar interaction
\cite{villain:79,saslow:88}, but it remains an open question whether collinear order
can be destroyed with an arbitrarily small concentration of sufficiently strong
defect bonds, and what role is played by the spatial dimension
\cite{parker:88,vannimenus:89,gawiec:91a,kruger:02}.  Such models are potentially
relevant to the weakly doped copper-oxide parents of high-$T_c$ superconductors
\cite{aharony:88a,ching:90,kruger:02}, while periodically frustrated {\em XY\/}
systems are models of Josephson junction arrays
\cite{hasley:85,vannimenus:89,reid:97,gupta:98}. For the strong disorder case of a
spin glass with symmetric bond distribution, the question of whether there is a
non-zero transition temperature (for either or both of the spin and chiral sectors)
in 3D {\em XY\/} and Heisenberg EA models is currently (again) hotly debated
\cite{lee:03a,yamamoto:04,kawamura:05a,campos:06,nakamura:06,campbell:07}.

For the SK model, the ordered phase is characterized by a breaking of the abstract
replica symmetry \cite{mezard:84}, but the nature of symmetry breaking in the short
range case is less clear. Starting out from an essential analogy with the behavior of
the ferromagnet, a phenomenological droplet scaling theory for the effective coupling
in spin glasses has been formulated
\cite{mcmillan:84a,bray:84,fisher:86,bray:87a,fisher:88}. The surface (free) energy
of droplet excitations of length $L$ is assumed to scale as $\Delta E \sim
L^{\theta_s}$, defining the spin stiffness exponent $\theta_s$. Put into the context
of a renormalization group (RG) framework \cite{bray:87a}, $\theta_s$ determines
whether a system scales to weak coupling ($\theta_s < 0$) such that droplets can
always destabilize the ordered phase, thus restricting its extension to zero
temperature, or scales towards strong coupling ($\theta_s > 0$) such that large-scale
excitations are sufficiently costly to allow for a stable spin-glass phase at finite
temperatures $0< T< T_\mathrm{SG}$. In the infinite-range SK model, the notion of
droplets is maldefined and, instead, it is found that excitations of arbitrary size
can be invoked at {\em constant\/} energy
\cite{parisi:79,parisi:80,parisi:80a,mezard:84}.  Consequently, if mean-field theory
applies to low-dimensional spin glasses, low-energy excitations are of rather
different nature implying, for instance, space filling domain walls
\cite{newman:03a}. Hence, an analysis of overlaps between ground and systematically
excited states can in principle allow to distinguish between the droplet and
mean-field pictures \cite{palassini:99,marinari:00}. A standard way of producing
excitations is a change of boundary conditions (BCs), e.g.\ from periodic to
antiperiodic, the energy difference between both configurations capturing the energy
of a relative domain wall \cite{banavar:82a,mcmillan:84a}.

For the 2D Ising spin glass, domain-wall energies scale with $\theta_s = -0.287(4)$
for Gaussian bond distribution \cite{hartmann:02a} and $\theta_s = 0$ for bimodal
couplings \cite{hartmann:01a}, and a true thermodynamic spin-glass phase occurs only
at zero temperature in both cases.  From the RG ansatz of Ref.\
\onlinecite{bray:87a}, $\theta_s$ determines the divergence of the correlation length
$\xi$ as $T \rightarrow 0$ via $\nu = -1/\theta_s$, which is found to be consistent
with finite-temperature Monte Carlo results for the Gaussian case
\cite{houdayer:04,katzgraber:04}. For the bimodal $\pm J$ distribution, on the other
hand, the formal $\nu = \infty$ has been interpreted as {\em exponential\/}
divergence \cite{katzgraber:05a} of $\xi$, but some evidence for an algebraic
singularity has also been presented recently \cite{jorg:06} (see also
Ref.~\onlinecite{fisch:06a}). For the case of continuous {\em XY\/} or Heisenberg
spins, the situation is complicated by the fact that the non-collinear spin structure
in the ordered phase allows for the distinction of proper and improper O($n$)
transformations, such that the continuous SO($n$) rotational symmetry is augmented by
a $\mathbb{Z}_2$ or Ising-like {\em chiral\/} symmetry related to the determinant of
the O($n$) rotation matrix \cite{villain:77a}. It has been argued \cite{kawamura:87a}
that these chiral degrees-of-freedom might decouple from the rotational ordering,
resulting in different transition temperatures (above the lower critical dimension)
or, at least, an additional chiral stiffness exponent $\theta_c$ (for
$T_\mathrm{SG}=0$). Such a scenario is --- after a long-winded debate --- now quite
well established in some periodically frustrated magnets
\cite{loison:00,ozeki:03,hasenbusch:05}, but remains fervently debated for the case
of spin glasses \cite{lee:03a,kawamura:05a,campos:06,campbell:07,kawamura:07}. For
the 2D {\em XY\/} spin glass, neither has $\theta_s$ been determined consistently,
with estimates ranging from \cite{jain:86a} $\theta_s = -1.0$ to
\cite{kosterlitz:99a} $\theta_s = -0.4$, nor has the question of a possible
spin-chirality decoupling been satisfactorily resolved
\cite{maucourt:98a,kosterlitz:99a}.

Methodologically, the above uncertainties regarding the {\em XY\/} spin glass are
related to various technical difficulties.  Firstly, finding ground states of
spin-glass model systems is a computationally hard problem, generically believed to
require an effort growing exponentially with system size \cite{hartmann:book}. The 2D
Ising spin glass on planar graphs is the only non-trivial exception to this rule,
being polynomial computationally \cite{bieche:80a}, and allowing for the rather
precise estimate of $\theta_s$ cited above. Thus, reliable spin-glass ground state
computations depend in general on rather elaborate and specifically tailored
approximation algorithms to treat reasonably sized systems. In this context, we use
here the recently proposed ``genetic embedded matching'' (GEM) technique
\cite{weigel:05f,weigel:06b} to investigate the zero-temperature properties of the 2D
{\em XY\/} spin-glass model.  For the Ising problem mentioned, due to the existence
of rather pronounced finite-size effects \cite{hartmann:03a} relatively large system
sizes turned out to be necessary for reliable estimates, e.g., of $\theta_s$. Below,
we will see how progress in accessing much larger system sizes using the GEM approach
now allows for investigations where finite-size effects seem for the first time
rather well controlled. Secondly, there is no full agreement as to how BCs should be
chosen in order to precisely excite either spin or chiral defects
\cite{kosterlitz:99a,uda:05}.  This is of particular concern here, given that strong
corrections to scaling depending on the chosen set of BCs have been observed for the
Ising case \cite{carter:02a,hartmann:02a}. Here, we systematically study a wide range
of BCs and resolve the boundary dependent corrections to scaling.

The rest of the paper is organized as follows. Section \ref{sec:algo} briefly
discusses the employed GEM technique and the technical set-up of the calculations. In
Sec.\ \ref{sec:gs_energy}, we discuss the asymptotic ground-state energy and the
spectrum of metastable states found in the GEM calculations, while Sec.\
\ref{sec:walls} reports on the properties of excitations induced by a change of BCs.
In particular, spin and chiral stiffness exponents and the associated fractal
dimensions of domain walls are determined. Some of these results already were
reported in part in Ref.\ \onlinecite{weigel:05f}. Section \ref{sec:overlaps}
contains our results for the scaling of overlap distributions, and
Sec.~\ref{sec:further} presents some results on the distribution of internal fields
and the disorder strength vs.\ disorder concentration phase diagram of the 2D
randomly frustrated {\em XY\/} magnet. Finally, Sec.\ \ref{sec:concl} contains our
conclusions.

\section{Algorithm and technical set-up\label{sec:algo}}

Zero-temperature properties might be inferred from the $T\rightarrow 0$ limit of
finite-temperature calculations such as Monte Carlo simulations. The latter, however,
suffer from severe slowing down as the $T=0$ critical point is approached for this
system, thus rendering the extrapolated results rather unreliable. Hence we chose,
instead, to determine ground states directly from non-equilibrium methods. Apart from
the 2D Ising EA model on planar graphs, such computations are believed (and in some
cases proven) to be {\em NP\/} hard\footnote{For practical purposes, this means that
  although the penalty of a solution (i.e., here the energy of the spin
  configuration) can be computed in polynomial time, it is believed that there exists
  no algorithm for surely finding the optimal solution in polynomial time.}
optimization problems \cite{hartmann:book}. To make progress with the 2D {\em XY\/}
model, we take inspiration from the approach making the 2D Ising EA model solvable
\cite{bieche:80a}. In particular, we {\em embed\/} Ising variables into the
continuous {\em XY\/} spins: with respect to a random, but fixed direction $\bm{r}$
in spin space, the O($n$) spins $\bm{S}_i$ of (\ref{eq:EA_model}) decompose as
$\bm{S}_i = \bm{S}_i^\parallel + \bm{S}_i^\perp = (\bm{S}_i\cdot\bm{r})\bm{r} +
\bm{S}_i^\perp$.  This results in a decomposition of ${\cal H}$ as ${\cal H} = {\cal
  H}^{r,\parallel} + {\cal H}^{r,\perp}$ with
\begin{equation}
  \label{eq:embedded_hamiltonian}
  {\cal H}^{r,\parallel} = -\sum_{\langle i,j\rangle}\tilde{J}^r_{ij}\,\epsilon_i^r
  \epsilon_j^r,
\end{equation}
where $\epsilon_i^r = \mathrm{sign}(\bm{S}_i\cdot\bm{r})$, and where the effective
couplings $\tilde{J}^r_{ij}$ are given by
\begin{equation}
  \label{eq:embedded_couplings}
  \tilde{J}_{ij}^r = J_{ij}|\bm{S}_i\cdot\bm{r}||\bm{S}_j\cdot\bm{r}|.
\end{equation}
If now the original O($n$) rotations of the spins are restricted to reflections along
the plane defined by $\bm{r}$, $\tilde{J}_{ij}^r$ as well as ${\cal H}^{r,\perp}$ are
easily seen to be invariant under such reflections of individual spins, whereas the
$\epsilon_i^r$ change sign. Hence, for a given $\bm{r}$, ${\cal H}^{r,\parallel}$ in
(\ref{eq:embedded_hamiltonian}) takes on the form of an Ising model with effective
Ising ``spins'' $\epsilon_i^r$.

The tractability of the 2D Ising ground-state problem in polynomial time comes about
via the transformation to a graph-theoretical formulation \cite{bieche:80a}. It rests
on a dualization of the model: define elementary plaquettes of the square lattices to
be frustrated if one or three of their surrounding bonds are antiferromagnetic
\cite{toulouse:77a}. For a given spin configuration, mark dual bonds of the square
lattice as ``active'' if the corresponding original bond is unsatisfied, i.e., if
$J_{ij} S_i S_j < 0$. Then, from the very definition of a frustrated plaquette, the
set of ``active'' (dual) bonds consists of individual chains which are either closed
loops or open chains connecting pairs of frustrated plaquettes, and the total energy
of ${\cal H}$ in (\ref{eq:EA_model}) is (up to a constant) identical to the total sum
of $J_{ij}$ along the ``active'' bonds. Since on a planar lattice closed loops can
always be removed by contraction, a ground state of the spin system corresponds to a
{\em minimum-weight perfect matching\/} of the frustrated plaquettes, i.e., a
configuration of the active bonds matching up all pairs of frustrated plaquettes
while minimizing the total weight $\sum J_{ij}$ of bonds contained. Such problems can
be solved in polynomial time by means of Edmonds' ``blossom algorithm''
\cite{edmonds:65a}, and, most importantly, the solution can be shown to always
correspond to a valid spin configuration for the case of planar graphs
\cite{bieche:80a}. As a result, application of (anti-)periodic BCs is restricted to
at most one direction to preserve planarity. See
Refs.~\onlinecite{bieche:80a,hartmann:book,weigel:06b} for a more detailed account of
the relation between the Ising ground state and graph-theoretical matching problems.

Using the decomposition (\ref{eq:embedded_hamiltonian}) for the O($n$) Hamiltonian
(\ref{eq:EA_model}), the blossom algorithm can be used to find exact ground states of
the embedded Ising spins $\epsilon_i^r$, corresponding to a reflection of some of the
continuous spins with respect to the plane defined by a randomly chosen direction
$\bm{r}$. A random sequential sampling of directions $\bm{r}$ is used to
statistically recover the full O($n$) symmetry. This {\em embedded matching\/}
procedure is a strictly downhill search which, however, due to its non-local nature
originating from the exact ground-state computation for the embedded Ising spins,
features far less metastable states than a zero-temperature quench with any local
dynamics \cite{walker:80}. Due to the dependence of the effective couplings
$\tilde{J}_{ij}^r$ of (\ref{eq:embedded_couplings}) on the configuration
$\{\bm{S}_i\}$, the embedded matching dynamics normally drives the system into a
(low-lying) metastable state instead of the true ground state.

To find true ground states, the above embedded matching must be supplemented by a
complementary protocol. We do so by inserting the embedded matching (i.e.,
minimization technique) as an optimization component (subroutine) into a specially
tailored {\em genetic algorithm\/} \cite{pal:96a,weigel:05f}. Here, a whole
population of candidate ground state configurations is being worked on, continuously
undergoing an improvement process mimicking natural evolution, which allows for a
very efficient exploration of the landscape of metastable states of the system:
\begin{itemize}
\item {\bf Crossover:\/} a random pair of spin configurations is selected as ``parent
  configurations'' and locally rigid domains or clusters are exchanged at random
  between them to produce ``offspring'' (see also Sec.~\ref{sec:spectrum}).
\item {\bf Mutation\/}: a small proportion of spins of either offspring are randomly
  perturbed.
\item {\bf Optimization\/}: both offspring are optimized using the embedded matching
  technique.
\item {\bf Selection\/}: each parent is replaced by the morphologically closer
  offspring in case the latter has a lower internal energy.
\end{itemize}
The most crucial step in this sequence is the ``crossover'' operation. Indeed,
special attention must be paid to the way configurations are combined in the
crossover to balance the need for ``genetic'' diversification with the preservation
of the high optimization on short length scales already achieved at intermediate
steps of the process.  Statistical analyzes show that the above combination of
techniques in this {\em genetic embedded matching\/} (GEM) approach allows for a
reliable determination of ground states for 2D {\em XY\/} spin glass samples of up to
about $30\times 30$ spins with current computing resources\footnote{Note that,
  although the embedded matching approach is polynomial, the {\em XY\/} ground-state
  problem itself is still {\em NP\/}-hard, such that, for instance, the population
  size in the genetic part needs to grow exponentially with system size at a constant
  success rate for finding ground states (see Ref.~\onlinecite{weigel:06b}).}. The
algorithmic details of the GEM technique are quite intricate. A presentation of those
details here would detract from our main agenda, which is to focus, and report, on
the {\it physical} properties of the ground state of the 2D {\em XY\/} spin glass. A
full account of the algorithmic details of the GEM technique can be found in
Ref.~\onlinecite{weigel:06b}.

For specificity, we assume a bimodal distribution of the quenched random couplings $J_{ij}$ of
(\ref{eq:EA_model}),
\begin{equation}
  \label{eq:coupling_distribution}
  P(J_{ij}) = (1-x)\,\delta(J_{ij}-J) + x\,\delta(J_{ij}+\lambda J), 
\end{equation}
where $\lambda$ denotes the relative strength of the antiferromagnetic impurity bonds
$J_{ij} = -\lambda J$ and $x$ their concentration. Unless stated otherwise, we
averaged with respect to the distribution (\ref{eq:coupling_distribution}) by taking
$5\,000$ independent disorder realizations and concentrated on the symmetric point
$x=0.5$, $\lambda = 1$, deep in the spin-glass regime of model (\ref{eq:EA_model})
with $P(J_{ij})$ from Eq.~(\ref{eq:coupling_distribution}). All computations
discussed here were performed on square lattices. Choosing the parameters of the GEM
runs to arrive at true ground states with high reliability, the run-times for single
ground-state computations on a single 2.8 GHz PentiumIV CPU are between a few seconds
for the smallest lattice sizes up to about 8 hours for a $28\times 28$ system. In
total, production of the data presented here consumed the equivalent of about $80$
years of single CPU time.

\section{Energetic properties\label{sec:gs_energy}}

\subsection{Asymptotic ground-state energy}

\begin{figure}[tb]
  \centering
  \includegraphics[clip=false,keepaspectratio=true,height=4.5cm]{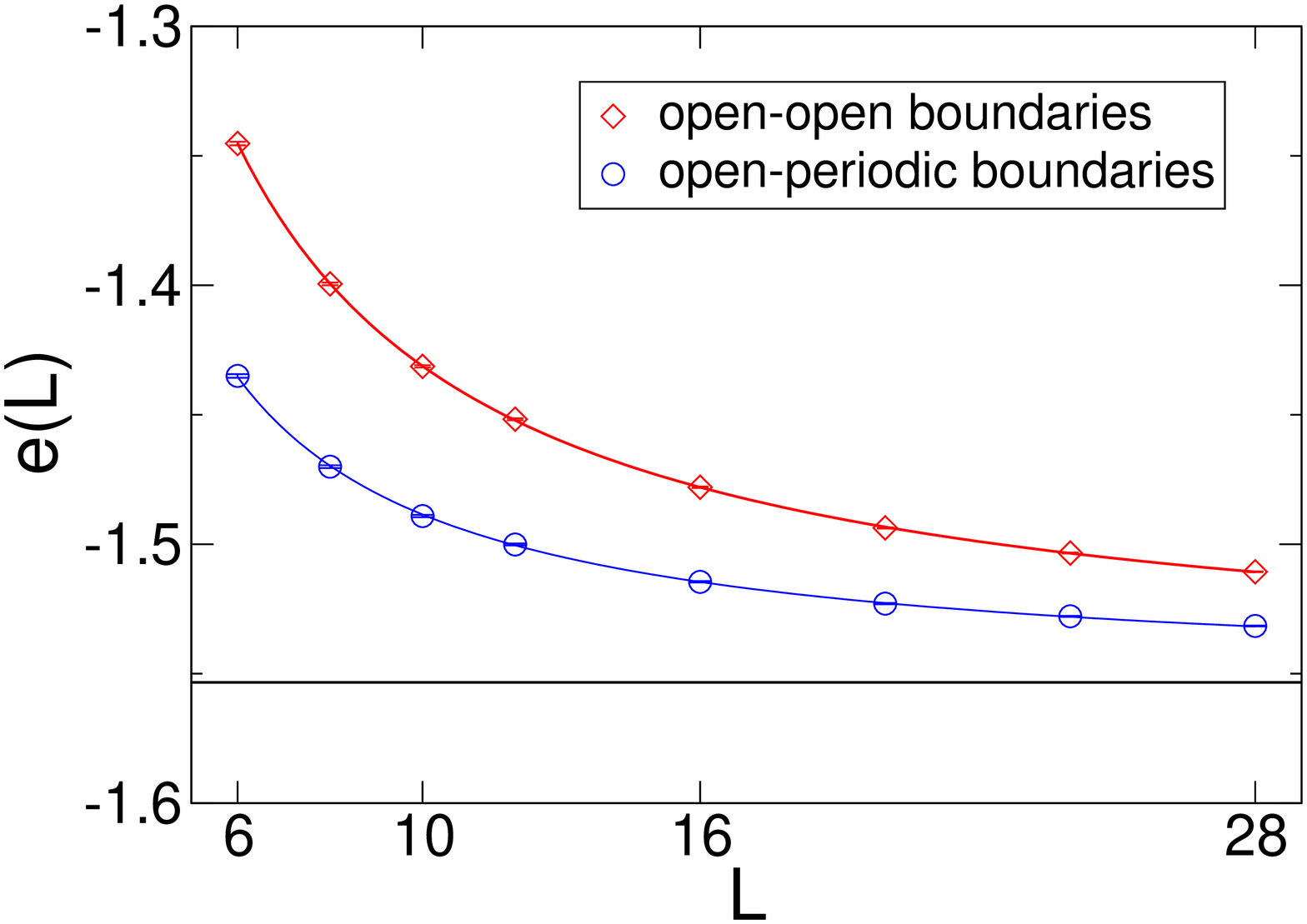}
  \includegraphics[clip=false,keepaspectratio=true,height=4.5cm]{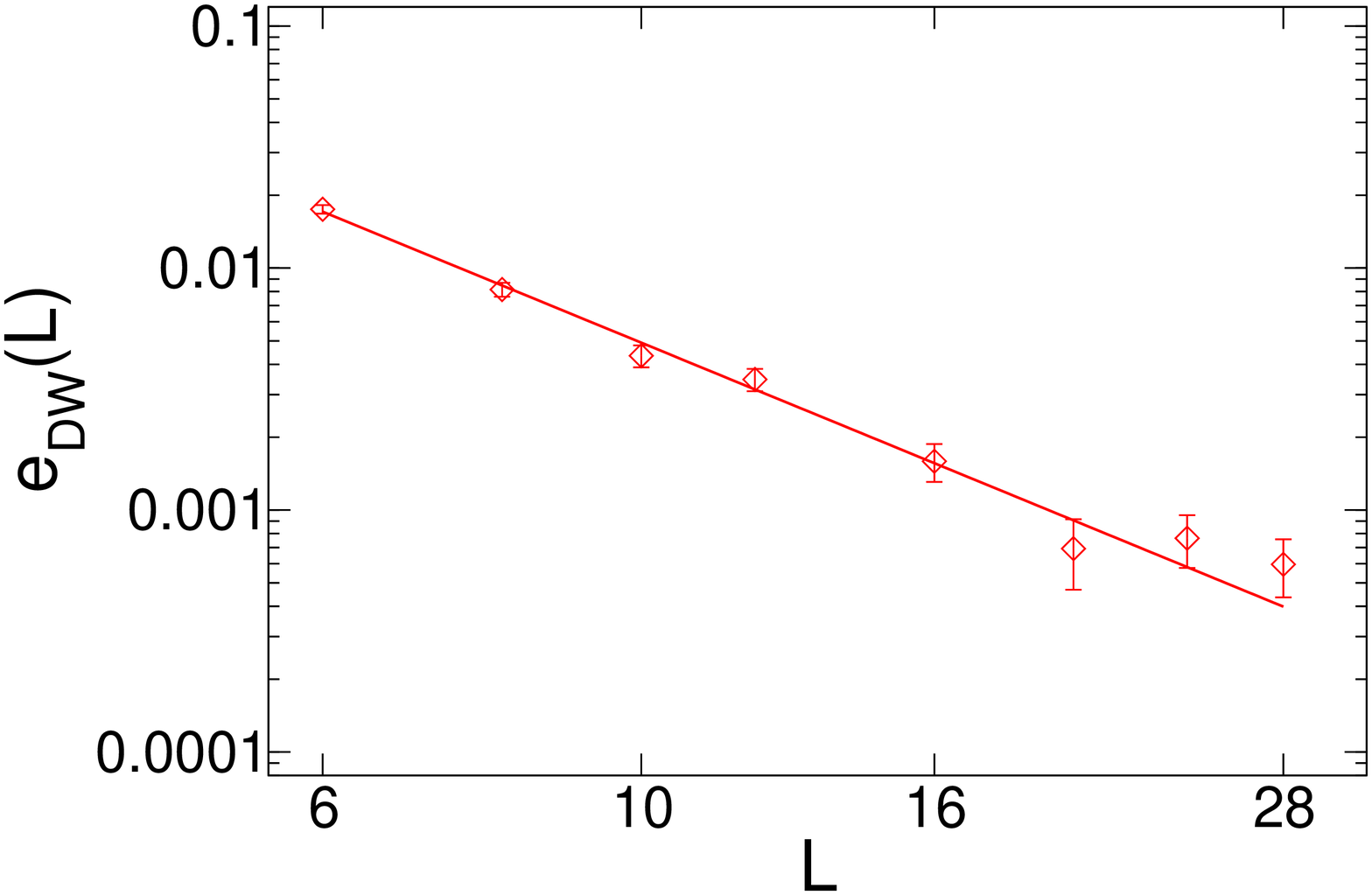}
  \caption
  {(Color online) Top: Average ground-state energy per site for open-open and
    open-periodic boundaries. The fits are described in the main text, and the
    horizontal line shows the estimated asymptotic value of $e_\infty = 1.55\,331$.
    Bottom: domain-wall related correction to the internal energy for open-periodic
    BCs, which we denote by $e_\mathrm{DW}(L)$, and which is given by
    $e_\mathrm{DW}(L) = e(L)-e_\infty-({\cal B}^\ast-e^\mathrm{bd}_\infty)/(2L)+{\cal
      B}^\ast/(4L^2)$.  Here, the open-boundary estimates $e_\infty = 1.55\,331$ and
    ${\cal B}^\ast = -0.375$ have been used. The line shows a fit of the form
    $e_\mathrm{DW}(L) = AL^{-(d-\theta_s)}$, yielding an estimate $\theta_s =
    -0.432(94)$.}
  \label{fig:energy_correction}
\end{figure}

To determine the asymptotic ground-state energy, we performed a finite-size scaling
analysis with systems of different BCs, namely open and open-periodic (as mentioned
above, fully periodic boundaries are not accessible to the method due to the
planarity constraint \cite{bieche:80a}). We used $L\times L$ systems with linear
dimensions $L=6$, 8, 10, 12, 16, 20, 24, and 28 and with $5\,000$ disorder replica
for each system size. We consider the average, size-dependent internal energy per
spin, $e(L) = \l E/N_0\r_J$, where $\l\cdot\r_J$ denotes a disorder average and $N_0$
is the number of lattice sites.  From generic finite-size scaling arguments
\cite{barber:domb}, at a $T=0$ critical point, $e(L)$ should scale as
\cite{hartmann:04a}
\begin{equation}
  \label{eq:energy_scaling}
  e(L)-e_\infty \sim L^{-(d-\theta_s)},
\end{equation}
where $d$ denotes the spatial dimension and $\theta_s = -1/\nu$ was assumed
\cite{bray:87a}. As usual, corrections to this form can be either non-analytic,
stemming from irrelevant operators, or analytic and proportional to $L^{-k}$, $k=1$,
2, $\ldots$, from non-linear scaling fields \cite{henkel:book}. From the
$\epsilon$-expansion of the Ising spin glass, the leading non-analytic correction is
(to first order in $\epsilon = 6-d$) found \cite{hartmann:04a} to be proportional to
$L^{\theta_s-6}$, asymptotically clearly smaller than the leading analytic correction
term. Assuming a similar behavior for continuous spins, we shall restrict ourselves
to analytical corrections here. Then, the most general form would be
\begin{equation}
  \label{eq:energy_corrections}
  e(L)-e_\infty = A L^{-(d-\theta_s)} +B L^{-1} + CL^{-2}+DL^{-3} + \ldots.
\end{equation}
We first consider the case of fully open boundaries. Noting that in this case the
analytical corrections can be at least partly attributed to the presence of edges and
corners (cf., e.g., Ref.~\onlinecite{privman:privman}), Campbell {\em et
  al}.\cite{hartmann:04a} take a slightly different approach and, instead of using
$e(L)$, consider the energy $e^\mathrm{bd}(L)/2$ per {\em bond\/} as the fundamental
property, making the following scaling ansatz,
\begin{equation}
  \label{eq:bond_energy_free}
  e^\mathrm{bd}(L) - e^\mathrm{bd}_\infty = {\cal B}^\ast/L + {\cal C}^\ast/L^2,
\end{equation}
with ${\cal B}^\ast$ representing edge effects and ${\cal C}^\ast$ accounting for
corner corrections. No term proportional to $L^{-(d-\theta_s)}$ is included in
Eq.~(\ref{eq:bond_energy_free}) arguing that free boundaries do not force any domain
walls into the system. Indeed, excellent agreement with the data is found under this
assumption for the case of the Ising spin glass \cite{hartmann:04a}.  Since $L\,e(L)
= (L-1)\,e^\mathrm{bd}(L)$, the parameters of Eqs.\ (\ref{eq:energy_corrections}) and
(\ref{eq:bond_energy_free}) are related as $e_\infty = e^\mathrm{bd}_\infty$, $B =
{\cal B}^\ast-e^\mathrm{bd}_\infty$, $C = {\cal C}^\ast-{\cal B}^\ast$, $D = -{\cal
  C}^\ast$, and $A = 0$. Note that this effectively reduces the number of fit
parameters of the form (\ref{eq:energy_corrections}) with $A=0$ from four to three,
the $1/L^3$ term merely being produced from the $1/L^2$ correction of
(\ref{eq:bond_energy_free}) by the transformation from bonds to sites. A fit of the
form (\ref{eq:energy_corrections}) with $A=0$ to our data for free boundaries yields
$e_\infty = -1.5520(14)$ with good quality-of-fit $Q=0.35$. Using ${\cal B}^\ast$ and
${\cal C}^\ast$ as free parameters, on the other hand, yields a very similar estimate
$e_\infty = -1.55\,331(56)$ with $Q = 0.48$. This last fit is shown together with the
data in the upper panel of Fig.~\ref{fig:energy_correction}. It is interesting to
note that the resulting parameter estimates ${\cal B}^\ast = -0.375(17)$ and ${\cal
  C}^\ast = 0.059(98)$ are in fact rather close to the values ${\cal B}^\ast =
-0.3205(9)$ and ${\cal C}^\ast = 0.042(3)$ found for the {\em Gaussian\/} 2D Ising
case and much further away from the estimates ${\cal B}^\ast = -0.5492(20)$ and
${\cal C}^\ast = 0.506(18)$ for the {\em bimodal\/} Ising spin glass
\cite{hartmann:04a}. This is consistent with the expectation that the continuity of
the {\em XY\/} spins effectively smoothes out the discrete bimodal distribution of
couplings $J_{ij}$ as far as energetic defects are concerned.

For (mixed) open-periodic boundaries there is no reason to exclude an
$L^{-(d-\theta_s)}$ term related to domain-wall trapping. In the scaling ansatz
(\ref{eq:bond_energy_free}) for the bond energies, the $1/L^2$ correction should be
omitted since the system has no corners, and the edge contribution should be cut in
half. Since now $2L\,e(L) = (2L-1)\,e^\mathrm{bd}(L)$, one identifies $B = ({\cal
  B}^\ast-e^\mathrm{bd}_\infty)/2$ and $C = -{\cal B}^\ast/4$. The statistical
precision of the data is not sufficient to perform an unrestricted fit of the general
form (\ref{eq:energy_corrections}). Fixing $C=D=0$ with the four unrestricted
parameters $e_\infty$, $A$, $\theta_s$ and $B$ remaining, we arrive at $e_\infty =
-1.5525(13)$, $\theta_s=-0.49(69)$ and $Q=0.35$.  Fixing only $D=0$, but reducing the
number of parameters to four again by using the parametrization with ${\cal B}^\ast$
and ${\cal C}^\ast$ yields a rather similar estimate $e_\infty = -1.5529(13)$,
$\theta_s=-0.59(77)$ and $Q=0.40$. Using the estimates $e_\infty = -1.55\,331(56)$
and ${\cal B}^\ast = -0.375(17)$ from the open-boundary case to reduce the number of
parameters to two, on the other hand, yields a good fit ($Q = 0.38$) with a much more
precise estimate $\theta_s = -0.432(94)$ (note, however, that the quoted statistical
error does not take the uncertainty in the fixed parameters $e_\infty$ and ${\cal
  B}^\ast$ into account).  This fit is displayed together with the data in the upper
panel of Fig.~\ref{fig:energy_correction}. The lower panel shows the domain-wall
related part $e_\mathrm{DW}(L)$ of the energy correction for the open-periodic case.
As is clearly seen, the domain-wall contribution is a very small correction to the
internal energy. The estimate for $\theta_s$ is compatible with the result found from
the domain-wall calculations for periodic BCs at fixed aspect-ratio $R = 1$ presented
below in Sec.~\ref{sec:walls}, but too large in modulus compared to our final
estimate from the aspect-ratio scaling. One interesting question is whether spin
and/or chiral defects should contribute to the domain-wall correction to the internal
energy. We will show below in Sec.~\ref{sec:walls} that the chiral defects are of
much lower absolute energy for the considered lattice sizes than the spin defects,
such that essentially only the spin defects are at play here, justifying post factum
the interpretation of $\theta_s$ as the spin-stiffness exponent. Our best estimate
for the asymptotic ground-state energy, $e_\infty = -1.55\,331(56)$, should be
compared with the value $e_\infty = -1.401\,97(2)$ of the bimodal 2D Ising spin glass
\cite{hartmann:04a} which is, perhaps surprisingly, only about $10\%$ higher. Hence,
the enhanced ability of the continuous model compared to the discrete Ising case to
locally adapt does not allow to substantially relieve the frustration imposed by the
bond disorder (the unfrustrated system yielding, of course, $e_\infty = -2$).

\subsection{Spectrum of metastable states\label{sec:spectrum}}

While a finely-tuned choice of parameters in the GEM approach guarantees high
reliability for the optimization to arrive at a ground state, less optimal settings
(in particular a reduction of the initial population size below a certain threshold)
result in runs converging to low-lying metastable states instead
\cite{weigel:05f,weigel:06b}. Additionally, and irrespective of the parameter choice,
a part of the spectrum of low-lying metastable states appears as intermediate
population members in the course of the optimization procedure. For a finite system
this spectrum appears discrete\footnote{Spin-wave excitations out of local minima do
  not occur in the set of these states because after each ``optimization'' step of
  the GEM the configuration is forced into the closest local minimum through a
  relaxational dynamics.} (see below), and one naturally wonders what the nature of
these excitations is.  Comparing the configurations of some of these states, it is
clearly seen that on short length scales the system is rigid within the space of
energy minima, i.e., it naturally decomposes into stiff {\em clusters\/} of solidary
spins whose individual moments have very similar orientations in all of the low-lying
metastable states, apart from relative, essentially exact O($n$) rotations of whole
clusters.  Indications of such rigid clusters have been found in previous studies of
Ising \cite{bieche:80a,hed:01} and continuous-spin systems
\cite{henley:84b,henley:84a,maucourt:98a}.  However, the genuine existence of such
clusters for continuous spins had so far not been as well established numerically as
we are able to achieve here with the GEM method. Note that such clusters are the
finite-energy analogue of the ``backbone'' or ``rigid lattice'' of sites with
constant relative spin orientation in all ground states observed in the $\pm J$ Ising
spin glass with a large ground-state degeneracy \cite{barahona:82a}. The GEM
approach, and ultimately, its ability to determine the ground states, heavily relies
on this cluster structure in the choice of operation for the crossover procedure
exchanging such rigid domains between the ``parent'' replica, cf.\ the discussion in
Sec.~\ref{sec:algo} and Ref.~\cite{weigel:06b}. As a result, the GEM algorithm
effectively operates directly on the space of metastable states instead of on the
whole of the continuous phase space \cite{weigel:06b}. Products of independent O($n$)
transformations of adjacent clusters (plus local excitations close to the boundaries
between clusters) correspond to the domain-wall or droplet excitations postulated by
the droplet theory of the spin glass phase \cite{fisher:86,bray:87a,fisher:88} and
directly observed in Ising spin glasses
\cite{fisher:88,hartmann:01a,hartmann:03a}. Finally, from the perspective of defect
theory, the continuous symmetry of the order parameter also allows for {\em
  topological\/} excitations, here, according to the homotopy groups of O($2$),
realized as domain-walls and vortex excitations \cite{toulouse:79a}. These can be
explicitly observed in configurations \cite{henley:84b,weigel:06b}, contributing to
the thus rather rich variety of excitations appearing in this system.

\begin{figure}[tb]
  \centering
  \includegraphics[clip=true,keepaspectratio=true,width=5.5cm]{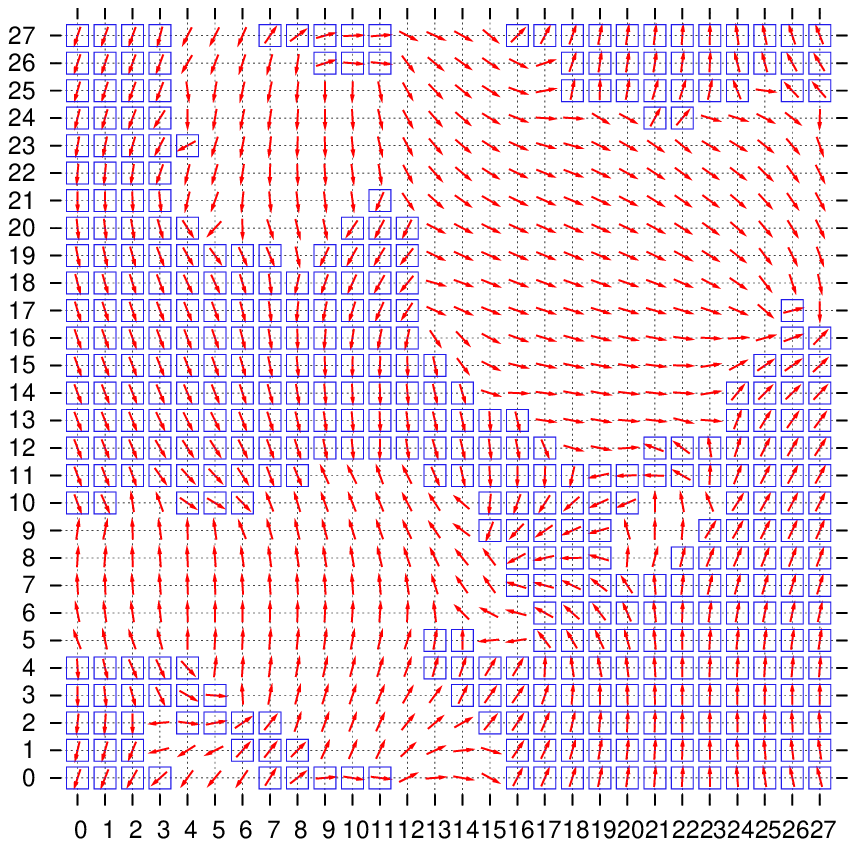}
  \includegraphics[clip=true,keepaspectratio=true,width=5.5cm]{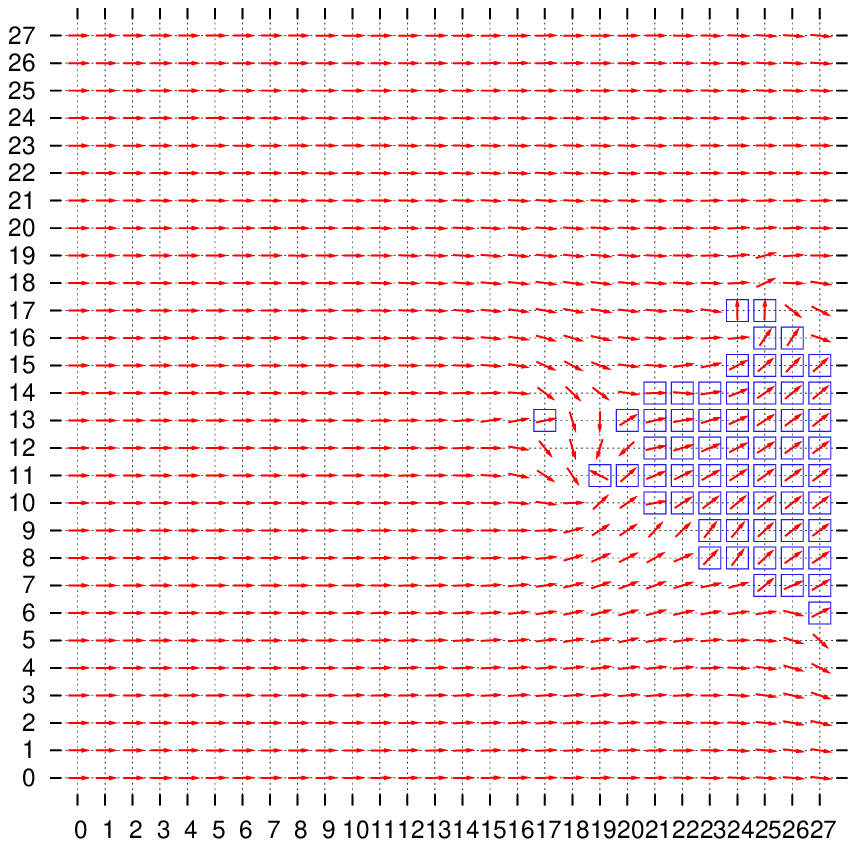}
  \caption
  {(Color online) Local rotation matrices between the ground and first excited states
    for two $28 \times 28$ disorder realizations with open-periodic boundaries
    (periodic boundaries in the vertical direction). The arrows indicate the rotation
    angles, and blue squares are drawn in the case of improper rotations, i.e.,
    negative determinant of the O($2$) matrix. See below in Sec.~\ref{sec:overlaps}
    for details about how these matrices are computed. The relative excitation
    energies are $\Delta E/E \approx 1.5\times 10^{-5}$ (upper configuration) and
    $\Delta E/E \approx 2.4\times 10^{-6}$ (lower configuration), respectively.}
  \label{fig:snapshots}
\end{figure}

Although a multitude of states with energies just slightly above the ground state
(with, e.g., $\Delta E/E$ as low as $\approx 10^{-6}$ for some $16\times 16$ systems)
is found, these energy differences are still far above the numerical resolution
limits, and the configurations can be shown for each case to be truly different (see,
for instance, the state differences depicted in Fig.~\ref{fig:snapshots}). Naturally,
each of these states has an $|\mathrm{O}(n)|$ degeneracy corresponding to a global
(proper or improper) rotation of all spins.  Conversely, however, we checked for
additional non-trivial degeneracies and found that statistically independent GEM runs
always converged to an O($n$) rotated copy of the same state. Hence, the ground state
of the system appears unique modulo the global symmetry. Consequently, quantities
related to statistical fluctuations vanish at $T=0$ and $\eta$, the exponent of the
correlation function, is equal to zero \cite{bray:87a}. In terms of the clusters of
rigid spins observed, the unique ground state results from the fact that the rigid
clusters are not free, but their rigid O($n$) rotations have some non-vanishing
excitation energy from necessary local adaptations of boundary spins.

By design, the GEM approach does not explicitly find {\em all\/} states of a given
energy. In particular, as a non-equilibrium technique it does not find states with
probabilities corresponding to their respective Boltzmann weights. However, there is
no reason to believe that in all cases a single of possibly a multitude of degenerate
ground states would dominate so strongly that not a single degenerate state appears
in the thousands of independent runs undertaken.  In contrast, the combined binary
symmetry of the bimodal Ising spin glass results in a massive degeneracy of the
ground state \cite{bieche:80a}, such that already a typical $10\times 10$ system has
between $10^4$ and $10^6$ ground states \cite{landry:02}.  Monte Carlo simulations
\cite{jain:86a,ray:92a} and Migdal-Kadanoff RG calculations \cite{gingras:92b} have
led to speculations that the pairing of bimodal couplings with the discrete chiral
symmetry might give rise to a similar situation in the bimodal {\em XY\/} spin-glass
model considered here. Certainly, if calculations such as finite-temperature Monte
Carlo simulations are sensitive to a (possibly narrow) {\em range\/} of energies, the
observed dense spectrum of states might effectively appear as quasi-degenerate,
resulting in an erroneously non-zero estimate of $\eta$ from such simulations
\cite{jain:86a,ray:92a}. Also, one might speculate about degeneracies in excited
states which we have not investigated here, but might have shown up in some of the
previous studies.

\begin{figure}[tb]
  \centering
  \includegraphics[clip=true,keepaspectratio=true,height=4.5cm]{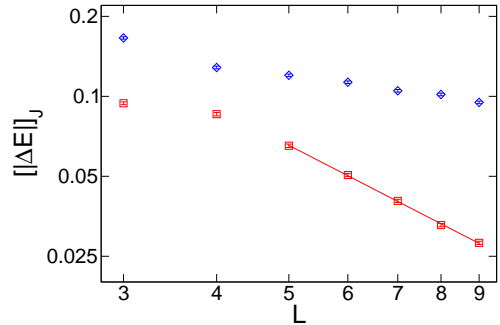}
  \caption
  {(Color online) Scaling of the energy gap between ground state and lowest excited
    metastable state as estimated from the GEM approach for $L\times 6L$ systems with
    open-periodic BCs. The line shows a fit of the form $\l\Delta E_\mathrm{gap}\r_J
    \sim L^{\theta_\mathrm{gap}}$ to the data. For comparison, the upper data set
    shows the average energy of domain walls induced by a change from periodic to
    antiperiodic BCs.}
  \label{fig:energy_gaps}
\end{figure}

The low-lying metastable states appearing in the process of a GEM run might allow to
learn something about the spectral properties of the system. Assuming a unique
leading exponent in the scaling of defect energies with length scale, it is
interesting to see how the lowest-energy excitations corresponding to the gap relate
to the domain-wall or droplet excitations that are crucial for the understanding of
the spin-glass phase. In Fig.~\ref{fig:energy_gaps} we show the gap energies for a
series of $L\times 6L$ systems with open-periodic BCs (the non-trivial aspect ratio
is chosen to reduce corrections to scaling, see the discussion below in
Sec.~\ref{sec:walls}). A fit of the functional form $\l\Delta E_\mathrm{gap}\r_J \sim
L^{\theta_\mathrm{gap}}$ to the data for $L\ge 5$ yields an estimate of
$\theta_\mathrm{gap} = -1.447(31)$. Thus, the energy gap closes rather rapidly,
leading to the expected quasi-continuous spectrum of metastable states in the
thermodynamic limit.  Also, $\theta_\mathrm{gap}$ is clearly far more negative than
the estimate $\theta_s \approx -0.3$ of the spin-stiffness exponent found below in
Sec.~\ref{sec:walls} from the domain-wall energy scaling. In fact, the energies of
domain walls induced by a change from periodic to antiperiodic boundaries,
additionally shown in Fig.~\ref{fig:energy_gaps} for comparison, are much larger than
the gap energies for the whole range of system sizes considered. It is thus clear (at
least as far as the range of available finite lattice sizes allows to judge) that the
excitations reflected in the gap scaling are {\em not\/} in general extended defects
of the domain-wall type. If, as argued here, metastable states are related to the
ground state by rigid O($n$) rotations of one or several clusters (plus certain local
adaptations at the cluster boundaries), it appears that the gapped excitations
correspond to rotations of single (or few) such clusters, resulting in a lower excess
energy than observed from the larger number of clusters that recombine to form a
domain-wall defect. This view is corroborated by the snapshots of defects
corresponding to the differences between ground and first excited states depicted in
Fig.~\ref{fig:snapshots}. Besides the extended defects with sizes of order $L$ which
might as well be incurred by a change of BCs (upper panel in
Fig.~\ref{fig:snapshots}), in many cases much more localized defects, corresponding
to single rigidly O($n$) rotated clusters, are the lowest excitations (lower panel in
Fig.~\ref{fig:snapshots}). These two situations (represented by the upper and lower
panels of Fig~\ref{fig:snapshots}) allow us to interpret the rapid drop of gap
energies indicated by the estimate of $\theta_\mathrm{gap}$ as the effect of an
additional implicit condition of minimum energy. For a given disorder configuration,
$\Delta E_\mathrm{gap}$ is not the average energy cost incurred by the O($n$)
rotation of a single cluster, but the minimum cost of all possible such rotations of
clusters.  Finally, it should be noted that $\Delta E_\mathrm{gap}$ is naturally not
confined to either spin or chiral \cite{villain:77a,kawamura:87a} sectors, such that
depending on the disorder configuration spin or chiral defects make up the gapped
excitation.

\section{Domain walls\label{sec:walls}}

For a systematic determination of the spin and chiral stiffness exponents $\theta_s$
and $\theta_c$, respectively, a controlled and direct insertion of the corresponding
defects into the system is needed.  This is most conveniently achieved by using an
appropriate choice of BCs.  Droplet excitations could be created by similar means
\cite{kawashima:00,hartmann:03a}, but we have not explored this yet.

\subsection{Choice and implementation of boundary conditions}

For the Ising spin glass, defect energies are carried by domain walls or droplet
surfaces which are the sharply localized boundaries of regions of inverted spins. For
continuous symmetry, the notion of defects is more subtle since the spin direction
varies smoothly, whereas chiralities are of the Ising type with localized boundaries
(cf.\ Fig.~\ref{fig:snapshots}). This is also seen in the types of topological
defects allowed by the O($n$) symmetry, namely plain domain walls for the Ising case,
but chiral domain walls {\em and\/} vortices for {\em XY\/} spins
\cite{toulouse:79a,henley:84a,henley:84b}. The classic approach of measuring defect
energies has been to analyze the difference between the energies $E_\mathrm{P}$ and
$E_\mathrm{AP}$ of systems with periodic (P) and antiperiodic (AP) BCs
\cite{banavar:82a}. In contrast to a ferromagnet, however, both P and AP boundaries
are in general frustrating for a spin glass such that, compared to the thermodynamic
state, in a finite system domain walls or other defects are forced into the system.
It is therefore not quite obvious what kind of excitation energy the difference
$\Delta E_\mathrm{P/AP}$ actually corresponds to. For the Ising case, the difference
or sum of two domain-wall excitations is still essentially a linear excitation of
length $L$, so one might expect to observe the scaling $|\Delta E|_\mathrm{P/AP} \sim
L^{\theta_s}$ predicted by droplet theory \cite{fisher:86,bray:87a}, which in fact
turns out to work reasonably well for sufficiently large system sizes
\cite{kawashima:03a}. The continuous case with spin and chiral variables is less
straightforward: the mere periodicity of periodic or antiperiodic BCs alone might
force either spin or chiral excitations into the system, such that the interpretation
of the difference $|\Delta E|_\mathrm{P/AP}$ is rather unclear. While previously only
P/AP boundaries (and the corresponding ``reflective'' boundaries for chiral defects)
had been considered \cite{morris:86a,kawamura:91,maucourt:98a}, Kosterlitz and Akino
\cite{kosterlitz:99a} suggested to alleviate the problem of trapped defects by
additionally optimizing over a global twist angle along the boundary and comparing
the resulting configuration to one with a relatively $\pi$ rotated or reflective
boundary (``optimum twist'' method). While such extra optimization over a global
twist minimizes the frustrating effect of periodicity, it does not remove it, and
hence spin and chiral defects remain incompletely disentangled.

\begin{figure}[tb]
  \centering
  \includegraphics[clip=true,keepaspectratio=true,height=4.5cm]{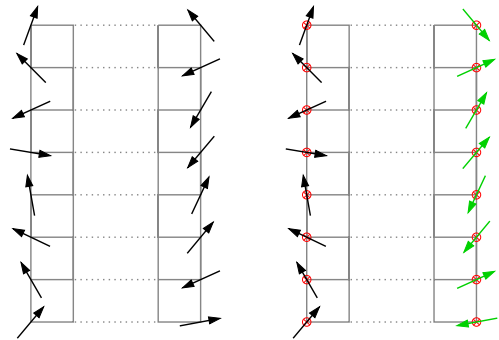}
  \caption
  {(Color online) Open and domain-wall BCs. A ground state for open boundaries is
    determined, resulting in some configuration of the boundary spins (left). For
    spin domain-wall boundaries, the spins along the right boundary are then rotated
    by $\pi$ and all boundary spins are held fixed for a second ground-state
    computation (right).  In the actual implementation used, instead of locking the
    boundary spins, only their relative orientation is fixed.}
  \label{fig:ODW_boundaries}
\end{figure}

A less ambiguous setup results from comparing the unrestricted, open-boundary system
with the case of an explicitly inserted defect. We refer to this approach as open (O)
and domain-wall (DW) BCs \cite{hartmann:01a}. The DW boundaries could be realized by
fixing the spins along one of the two boundary seams under consideration in the
positions of the O ground state and the spins of the other seam in a configuration
rotated by $\pi$ compared to the O state to induce a spin domain wall, cf.\ the
sketch in Fig.~\ref{fig:ODW_boundaries}. A fixing of spins is not possible within the
GEM approach, however, since the spins do not have a direct representation in the
dual matching problem.  Instead, we link the boundary seams periodically with strong
``anisotropic bonds'' guaranteeing a fixed relative orientation of spins. In the
embedded matching approach, for a pair $(\bm{S}_\mu, \bm{S}_\nu)$ of spins linked across
the boundary, we choose an effective coupling
\begin{equation}
  \label{eq:DW_embedded}
  \tilde{J}_{\mu\nu}^r = J_\mathrm{strong} \hat{\epsilon}_\mu^r\,\hat{\epsilon}_\nu^r
\end{equation}
instead of the general embedded coupling (\ref{eq:embedded_couplings}). Here,
$\hat{\epsilon}_\mu^r = \mathrm{sign}(\hat{\bm{S}}_\mu'\cdot\bm{r})$, $\hat{\bm{S}}_\mu'
= {\cal R}\hat{\bm{S}}_\mu$. The pair $(\hat{\bm{S}}_\mu, \hat{\bm{S}}_\nu)$ denotes the
intended final relative orientation of $(\bm{S}_\mu,
\bm{S}_\nu)$, i.e., in terms of the angles $\theta$, $\bm{S} = (\cos\theta, \sin\theta)$,
we have
\begin{equation}
  \label{eq:DW_angles}
  \begin{array}{rcl@{\hspace{0.4cm}}l}
    \hat{\theta_\mu}-\hat{\theta_\nu} & = & \theta_\mu-\theta_\nu+\pi & \mbox{for a spin DW},\\
     \hat{\theta_\mu}-\hat{\theta_\nu} & = & \theta_\mu+\theta_\nu & \mbox{for a chiral DW}.
  \end{array}
\end{equation}
Here, ${\cal R}$ denotes a rigid rotation of the pair $(\hat{\bm{S}}_\mu,\hat{\bm{S}}_\nu)$
such that $\hat{\bm{S}}_\nu = \bm{S}_\nu$. In this way, any relative orientation
$(\hat{\bm{S}}_\mu, \hat{\bm{S}}_\nu)$ can be forced upon the system by choosing a
(non-random) $J_\mathrm{strong}\gg \mathrm{max}\,|J_{ij}|$, such that a
$J_\mathrm{strong}$ bond is never broken in an embedded matching run (for our
computations, we chose $J_\mathrm{strong} = 10^5$). The resulting excess contribution
of the $J_\mathrm{strong}$ bonds is, of course, not counted in the final energy. For
snapshots of excitations incurred by these O/DW spin and chiral boundaries, see
Ref.~\onlinecite{weigel:06}. The prescription (\ref{eq:DW_angles}) for the chiral DW
boundaries corresponds to a fixed axis of reflection independent of the disorder
configuration (depending on the choice of origin for the angles $\theta_i$). As an
alternative, we have also considered a minimization procedure for the choice of the
reflection axis. In particular, we chose the reflection axis perpendicular to the
average direction of the boundary spins in order to minimize the boundary
perturbation. We found very similar results to the case of the prescription of
Eq.~(\ref{eq:DW_angles}), such that it appears that this ambiguity does not have any
bearing on the resulting value of the chiral stiffness exponent.  For completeness,
we also considered the standard P/AP BCs pair as well as random (R) and anti-random
(AR) BCs pair, where the spins are fixed in random relative orientations using the
procedure described above for the R computation and in relatively $\pi$ rotated
orientations for the computations with AR BCs.

\subsection{Spin and chiral stiffness}

\begin{figure}[tb]
  \centering
  \includegraphics[clip=true,keepaspectratio=true,height=4.5cm]{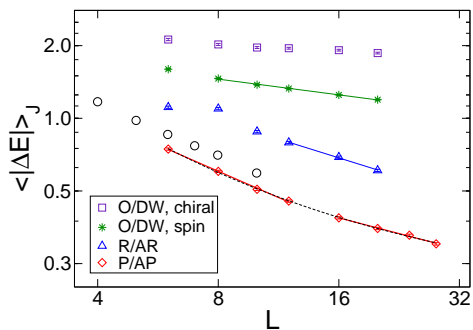}
  \caption
  {(Color online) Average defect energies $\l|\Delta E|\r_J$ for four sets of BCs on
    square lattices as a function of system size $L$. The straight lines are fits of
    the form $\l|\Delta E|\r_J\sim L^\theta$ to the data. The open black circles show
    the ``random twist'' result of Ref.~\onlinecite{kosterlitz:99a} for comparison.
    The dashed line shows a fit of the form (\ref{eq:dw_corrections}) with $C=D=0$ to
    the P/AP data. Some data sets have been shifted vertically for better
    distinction.}
  \label{fig:dw_scaling}
\end{figure}

We have computed defect energies for $L=6$ up to $L=28$ systems with P/AP, R/AR and
O/DW spin and chiral combinations of BCs. The results are collected in
Fig.~\ref{fig:dw_scaling} together with fits of the form $\l|\Delta E|\r_J \sim
L^{\theta}$ to the data. Alternatively, one might consider the scaling of the width
$\sigma(E) = \sqrt{\l(\Delta E-\l\Delta E\r_J)^2\r_J}$, which leads to very similar
results, and the final estimates of this approach will be discussed below. For the
P/AP combination, a pronounced crossover is observed from the sizes $L\le 12$ with
$\theta_s = -0.724(21)$ to $\theta_s = -0.433(26)$ for $L\ge 16$.  $L=12$ was about
the maximum size considered in previous studies \cite{maucourt:98a,kosterlitz:99a},
and our estimate for $L\le12$ is indeed consistent with these previous results as
illustrated by the P/AP data of Ref.~\onlinecite{kosterlitz:99a} shown for
comparison. The result for $L \ge 16$, on the other hand, is much smaller in modulus
and close to the value of $\theta_s \approx -0.4$ resulting from the ``optimum
twist'' prescription of Ref.~\onlinecite{kosterlitz:99a}, indicating that it indeed
reduces finite-size corrections to scaling. We note that the apparent crossover
length $L\approx 12$ is comparable to the length below which no metastability
occurs and the system behaves like a spherical spin glass \cite{morris:86a,lee:05a}.
The scaling of defect energies for the R/AR combination of boundaries behaves quite
similar to the P/AP case (as they are still periodic in nature), resulting in
$\theta_s = -0.519(30)$ for somewhat smaller systems. The O/DW setup, on the other
hand, not hampered by the periodicity effect, yields an even less negative estimate
$\theta_s = -0.207(12)$. Finally, for the chiral defects, we find $\theta_c =
-0.090(23)$.

As is seen from the variation of estimates with system size as well as between BCs,
corrections to scaling are strong. A very similar situation occurs for the Ising spin
glass \cite{hartmann:02a}. On general grounds \cite{barber:domb}, we expect analytic
and non-analytic finite-size scaling corrections,
\begin{equation}
  \label{eq:dw_corrections}
  \l|\Delta E|\r_J(L) = AL^\theta(1+BL^{-\omega})+C/L+D/L^2+\cdots. 
\end{equation}
In contrast to the scaling (\ref{eq:energy_corrections}) of the internal energy,
where the edge and corner effects were important, the dominant term is here given by
the non-analytic correction which is connected to interactions as well as
self-interactions of domain walls \cite{hartmann:03a}. Due to the limited precision
of the data, we have to set $C=D=0$. We find that only for the P/AP case are
corrections pronounced enough and a sufficient number of system sizes is available
for a stable fit that includes a non-analytic $L^{-\omega}$ correction term. For this
case (including all data points $L=6,\ldots,28$), we arrive at $\theta_s = -0.18(27)$,
$\omega = 1.21(28)$ with $Q = 0.43$ (cf.\ the dashed black curve in
Fig.~\ref{fig:dw_scaling}), whereas a fit including all data without the correction
term (i.e., additionally setting $B=0$) yields an unacceptable quality-of-fit
$Q=6\times 10^{-16}$. Hence, including corrections to scaling seems, in principle,
able to reconcile the data from different sets of boundaries. However, this can only
be done at the expense of mounting statistical uncertainties which cannot be afforded
for useful estimates with the available level of accuracy and range of system sizes.

\begin{figure}[tb]
  \centering
  \includegraphics[clip=true,keepaspectratio=true,height=4.5cm]{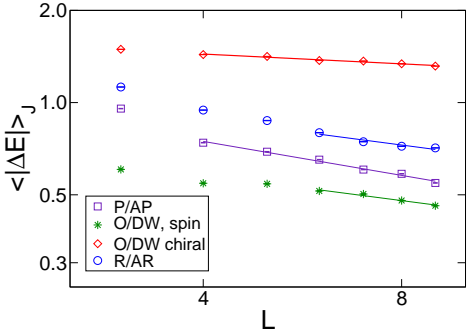}
  \caption
  {(Color online) Scaling of average defect energies $\l|\Delta E|\r_J$ for four sets
    of BCs at aspect ratio $R=6$. The straight lines are fits of the form $\l|\Delta
    E|\r_J\sim L^\theta$ to the data. Data sets have been shifted vertically for
    better distinction.}
  \label{fig:dw_energy_r6}
\end{figure}

\begin{figure}[tb]
  \centering
  \includegraphics[clip=true,keepaspectratio=true,height=4.5cm]{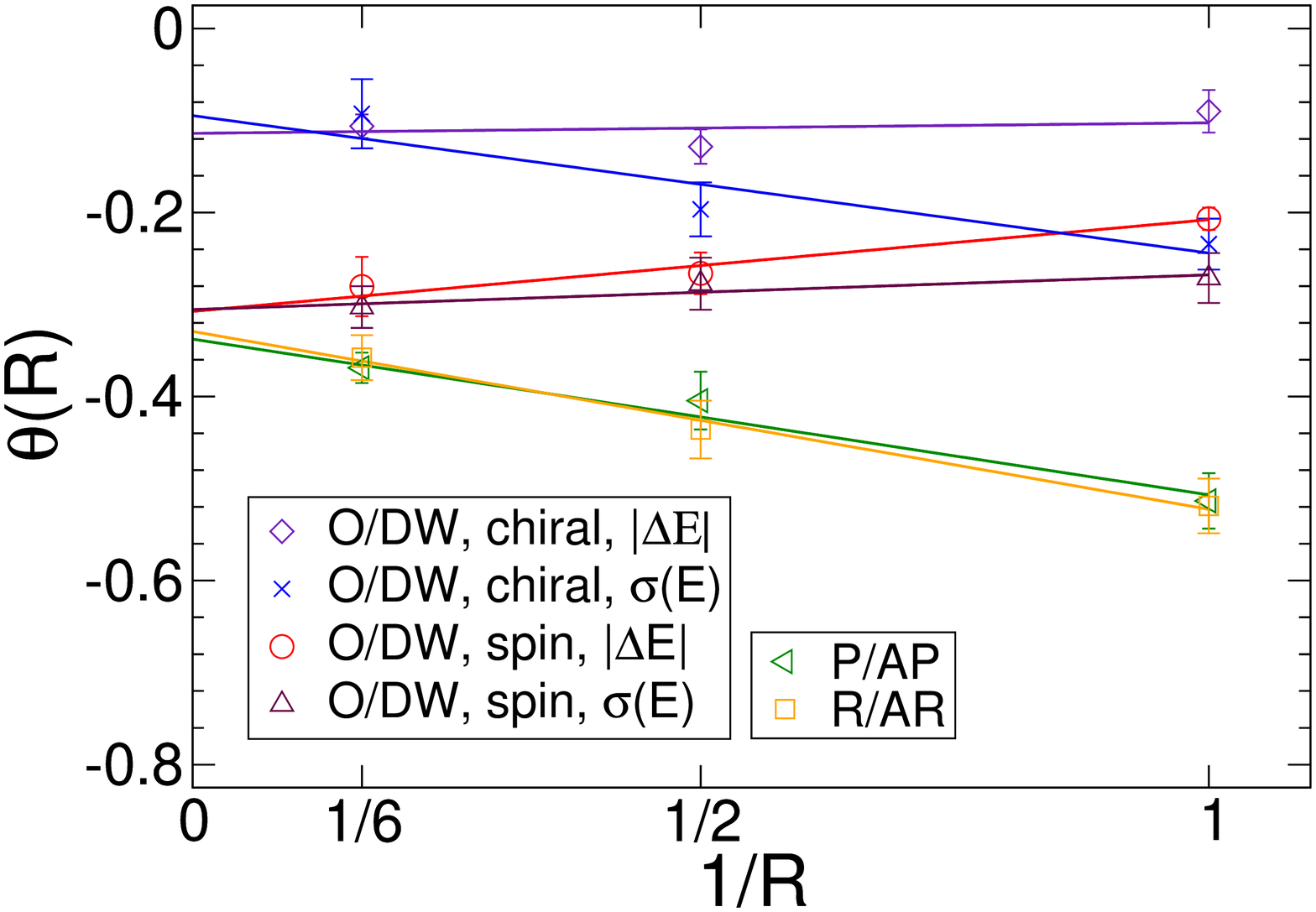}
  \caption
  {(Color online) Dependence of the stiffness exponents $\theta_s$ and $\theta_c$ on
    the aspect ratio $R$ of the systems for different pairs of BCs. The solid lines
    show fits of the form (\ref{eq:ARS_approach}).}
  \label{fig:ASR_scaling}
\end{figure}

As an alternative systematic treatment of scaling corrections for domain-wall
calculations, it has been suggested to consider $L\times M$ systems with variable
aspect ratios $R\equiv M/L$ (with the change of BCs occurring along the edges of
length $L$) \cite{carter:02a}. In general, scaling depends on the aspect ratio
\cite{barber:domb}, so we can write
\begin{equation}
  \label{eq:ARS_basic}
  \l|\Delta E|\r_J(L) = L^\theta F(R),
\end{equation}
with some scaling function $F(R)$. In the extreme case of very elongated systems,
$R\rightarrow\infty$, one should approach the behavior of the one-dimensional (1D)
system. For the 1D Ising model, it is found that $\l|\Delta E|\r_J(L) \sim 1/M$
independent of BCs \cite{bray:87a}. For the {\em XY\/} symmetry, the situation is
more involved due to the simultaneous presence of spin and chiral variables. It
appears that there are distinct exponents $\theta = (d-2) = -1$ related to spin waves
and $\theta \approx -1.9$ related to chiral excitations for the (quasi)
one-dimensional {\em XY\/} spin glass \cite{ney-nifle:93a,uda:05}. We are not aware,
however, of any tests of universality with respect to different BCs probing the same
excitations in this model. However, it is clear that the influence of the boundaries
must diminish for $R\gg 1$ as the average distance to the boundaries with variable
conditions increases. To test this hypothesis, we performed additional GEM
computations for systems with aspect ratios $R=2$ and $R=6$ with sizes
$L=4,6,\ldots,16$ ($R=2$) resp.\ $L=3,4,\ldots,9$ ($R=6$) and $5\,000$ disorder
replica per lattice. In Fig.\ \ref{fig:dw_energy_r6} we present the results for
domain-wall energies as a function of system size $L$ at aspect ratio $R=6$. As is
apparent, the slopes of the data for boundaries probing spin excitations, i.e., P/AP,
R/AR and O/DW spin boundary conditions, are quite compatible with each other
asymptotically, whereas energies of chiral domain walls decay considerably slower.
If, as it hence appears, there is independence of BCs as $R\rightarrow \infty$, one
expects scaling corrections depending on boundaries to gradually disappear as more
and more elongated systems are being considered. For the Ising case, it was observed
that to good approximation this limit is attained as \cite{carter:02a,hartmann:02a}
\begin{equation}
  \label{eq:ARS_approach}
  \theta(R) = \theta(R\!=\!\infty)+A_R/R.
\end{equation}
Our estimates of $\theta(R)$ for the data at aspect ratios $R=1$, $2$ and $6$ for the
different sets of boundary conditions are presented in Fig.~\ref{fig:ASR_scaling}
together with fits of the form (\ref{eq:ARS_approach}) to the data, also including
the alternative scaling of the width $\sigma(E)$. For each aspect ratio, corrections
at fixed $R$ were taken into account by omission of points from the small-$L$ side
for the fits of the form (\ref{eq:dw_corrections}) instead of including any
correction terms (i.e., we set $B=C=D=0$), cf.\ the fits presented in Fig.\
\ref{fig:dw_energy_r6}. In particular, data points were successively omitted from the
small-$L$ side until a satisfactory quality-of-fit $Q$ was achieved. We indeed find a
clear convergence of the P/AP, R/AR and O/DW spin stiffness exponents as more
elongated systems are considered, indicating independence of BCs in this limit. We
collect the extrapolated estimates in Table \ref{tab:ASR_results}, finding them all
well compatible within statistical errors. As our final estimate we cite the weighted
mean $\theta_s = -0.329(14)$ (we do not include the $\sigma(E)$ results in the
average since they are not statistically independent). The chiral stiffness exponent
$\theta_c$, on the other hand, seems to be almost independent of the aspect ratio (at
least as far as the scaling of the difference $|\Delta E|$ is concerned) with an
asymptotic value of $\theta_c(R=\infty)=-0.114(16)$, clearly different from the value
found for the spin stiffness, indicating spin-chirality decoupling in this model.

\subsection{Domain-wall fractal dimension\label{sec:fractal}}

On general grounds, the domain walls or droplet surfaces of the Ising spin glass have
been assumed to be meandering curves with a non-trivial fractal dimension $d_s$ (see
Refs.~\onlinecite{bray:87,fisher:88}). Measurements for the case of the 2D Ising spin
glass with Gaussian interactions indeed yield a non-generic value \cite{hartmann:02}
$d_s \approx 1.27$, and for the bimodal distribution quite similar estimates are
found \cite{weigel:06a,roma:07,melchert:07}. A similar value has also been found in
two-dimensional O($n$) spin glasses with additional random (off-diagonal) anisotropy
\cite{gingras:93}, which are expected to belong to the Ising spin-glass universality
class \cite{bray:82a}. The fractal dimension of domain walls is relevant to the
question of whether the mean-field scenario, which predicts space-filling domain
walls, i.e., $d_s = d$, applies to low dimensions \cite{newman:03a}. Consider the
link overlap,
\begin{equation}
  \label{eq:link_overlap}
  q_l^{AB} = \frac{1}{N_1} \sum_{\l i,j\r}
  (\bm{S}_i^{A}\cdot\bm{S}_j^{A}) (\bm{S}_i^{B}\cdot\bm{S}_j^{B}),
\end{equation}
where $N_1$ denotes the number of bonds of the lattice, and the superscripts $A$ and
$B$ stand for two replica of the system, here for the two BCs for a system with a
given realization of the random bonds.
For the Ising case, bonds crossed by the domain wall contribute $-1$ to $q_l^{AB}$ and all
the others $+1$, such that
\begin{equation}
  \label{eq:overlap_fractal_scaling}
  \langle 1-q_l^{AB}\rangle_J \sim L^{-(d-d_s)},
\end{equation}
which allows to determine $d_s$ numerically \cite{hartmann:02}. The authors of
Ref.~\onlinecite{katzgraber:02} used the same relation for continuous spins
(extrapolating from finite temperatures to $T=0$). Due to the continuous nature of
the spin response to the perturbation on the boundary, however, {\em all\/} spins
contribute here to the reduction of $q_l^{AB}$ and it is therefore almost inevitable
that $\l1-q_l^{AB}\r_J \sim L^0$, implying $d_s = d$. (It is likely that the small
difference $d-d_s$ found by the authors of Ref.~\onlinecite{katzgraber:02} is an
artefact of the extrapolation from finite temperatures.) Here, we indeed find, e.g.,
$d-d_s = 0.0010(14)$ for the P/AP case and similar results for the other boundary
conditions.  Due to the above argument, however, this should not be seen as evidence
for space-filling domain walls, but rather as an indicator for the inadequacy of the
method.

\begin{table}[tb]
  \centering
  \begin{ruledtabular}
    \begin{tabular}{ldd}
      Boundaries & \multicolumn{1}{c}{$|\Delta E|$} & 
      \multicolumn{1}{c}{$\sigma(E)$} \\ \colrule
      P/AP & -0.338(20) & -0.348(19) \\
      R/AR & -0.329(28) & -0.346(13) \\
      O/DW spin & -0.308(30) & -0.306(26) \\ \colrule
      O/DW chiral & -0.114(16) & -0.095(38)
    \end{tabular}
  \end{ruledtabular}
  \caption
  {
    Spin and chiral stiffness exponents $\theta(R=\infty)$ from different sets of
    BCs as extrapolated via aspect-ratio scaling according to
    (\ref{eq:ARS_approach}) for $|\Delta E|$ and the width $\sigma(E)$.
  }
  \label{tab:ASR_results}
\end{table}

As discussed above, domain walls in the system consist of the union of boundaries of
rigidly O($n$) rotated clusters. To reveal this structure in the ground states
identified, firstly one should get rid of the {\em global\/} symmetry by rigidly
O($n$) rotating one configuration such as to maximize the total overlap $\hat{q}^{AB}
= \sum_{\alpha,\beta} q_{\alpha\beta}^{AB} R_{\alpha\beta}^{AB}$, where
$R_{\alpha\beta}^{AB}$ denotes the corresponding global rotation matrix, and
\begin{equation}
  \label{eq:overlap_matrix}
  q_{\alpha\beta}^{AB} = \frac{1}{N_0} \sum_i S_{i\alpha}^{A} S_{i\beta}^{B},
\end{equation}
is the matrix of (site) overlaps. This is done by a singular value decomposition to
diagonalize $q_{\alpha\beta}^{AB}$, in which case $\hat{q}^{AB}$ is just the trace of
the resulting diagonal $q_{\alpha\beta}^{AB}$, see Ref.~\onlinecite{henley:84b}. We
refer to $\hat{q}^{AB}$ as the ``optimized'' overlap (or projection) in contrast to
the plain scalar overlap $q^{AB} = \sum_\alpha q_{\alpha\alpha}^{AB}$. Once this
global symmetry has been removed, one can attempt to determine {\em local\/} rotation
matrices to reveal rigid relative rotations of clusters of spins between the pair of
configurations considered. To this end, we consider a locally averaged overlap
matrix,
\begin{equation}
  \label{eq:averaged_overlap}
  q_{\alpha\beta}^{AB}(\bm{x}) = Z(\bm{x})^{-1}
  \sum_i w(\bm{x}_i-\bm{x})S_{i\alpha}^{A} S_{i\beta}^{B},
\end{equation}
where $Z(\bm{x}) = \sum_i w(\bm{x}_i-\bm{x})$, and $w(\bm{x}_i-\bm{x})$ is a rapidly
decaying weighting function. The optimal local rotation matrix
$R_{\alpha\beta}^{AB}(\bm{x})$ again follows from a singular value decomposition. The
averaging is necessary here since the comparison of a single pair of spins does not
allow to uniquely determine an O($2$) matrix. In particular, one could not decide
whether a proper or an improper rotation is involved. The weighting function
$w(\bm{x}_i-\bm{x})$ is chosen exponentially decaying and cut off after two or three
lattice spacings. The following results are independent of any fine-tuning of this
falloff. The snapshots of Fig.~\ref{fig:snapshots} show the angles and signs of the
thus determined rotation matrices $R_{\alpha\beta}^{AB}(\bm{x})$ (for snapshots of
domain walls induced by a change of BCs, see also Ref.~\onlinecite{weigel:06}).
Domain boundaries can be defined from $R_{\alpha\beta}^{AB}(\bm{x})$ separately for
the spin and chiral parts: for the latter any bond with different signs of the
determinant of $R_{\alpha\beta}^{AB}(\bm{x})$ at its two ends is crossed by a chiral
domain wall, and for the former any bond with a change of the rotation angle $\phi$
of $R_{\alpha\beta}^{AB}(\bm{x})$ by more than a threshold value $\phi_0$ is crossed
by a spin domain wall. To determine $d_s^\chi$ for the chiral case, we define a link
overlap analogous to (\ref{eq:link_overlap}),
\begin{equation}
  \label{eq:rotation_overlap}
  q_{l,\chi}^{AB} =  \frac{1}{N_1} \sum_{\l i,j\r}
  [\det \bm{R}^{AB}(\bm{x}_i)]\,[\det \bm{R}^{AB}(\bm{x}_j)],
\end{equation}
measuring the change in rotation signs along bonds. An analogous expression for
$q_{l,\phi}^{AB}$ is used for $d_s^\phi$ of the spin domain walls,
\begin{equation}
  \label{eq:rotation_overlap2}
  q_{l,\phi}^{AB} = \frac{1}{N_1} \sum_{\langle i,j\rangle} \theta_{\phi_0}(\bm{x}_i,\bm{x}_j),
\end{equation}
where $\theta_{\phi_0}(\bm{x}_i,\bm{x}_j) = -1$ if the angle between $\bm{S}_i$ and
$\bm{S}_j$ exceeds $\phi_0$ and $\det \bm{R}^{AB}(\bm{x}_i) = \det
\bm{R}^{AB}(\bm{x}_j)$, and $\theta_{\phi_0}(\bm{x}_i,\bm{x}_j) = +1$ otherwise.
Hence, bonds including a sign change of $\det \bm{R}^{AB}(\bm{x})$ are explicitly
excluded from the counting to optimally decouple spin and chiral measurements. Fits
of the form (\ref{eq:overlap_fractal_scaling}) are used to determine the
corresponding fractal dimensions.

\begin{figure}[tb]
  \centering
  \includegraphics[clip=false,keepaspectratio=true,height=4.5cm]{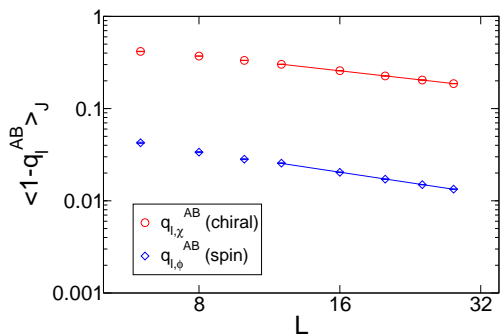}
  \caption
  {(Color online) Scaling of the chiral link overlap $q_{l,\chi}^{AB}$ of
    Eq.~(\ref{eq:rotation_overlap}) and the spin overlap $q_{l,\phi}^{AB}$ of
    Eq.~(\ref{eq:rotation_overlap2}) for $L\times L$ P/AP systems. The lines show
    fits of the functional form (\ref{eq:overlap_fractal_scaling}) to the data.}
  \label{fig:fractal_scaling}
\end{figure}

We consider the same sets of BCs and system sizes used for the determination of the
stiffness exponents above. Figure \ref{fig:fractal_scaling} shows the $L$ dependence
of the spin and chiral link overlaps $q_{l,\phi}^{AB}$ and $q_{l,\chi}^{AB}$,
respectively, for the case of P/AP BCs and $L\times L$ systems, revealing clear
scaling signals.  For the chiral overlap, a fit of the form
(\ref{eq:overlap_fractal_scaling}) to the data yields $d_s^\chi =1.4291(44)$. The
O/DW chiral boundaries presumably should show the clearest signal since direct visual
inspection of the configurations reveals that typically a single domain wall roughly
parallel to the variable boundary is excited.  For this case we arrive at $d_s^\chi
=1.394(12)$. The sizable, although not dramatic, difference for the two choices of
BCs shows that, as expected, still only partially controlled finite-size effects are
present. Unfortunately, aspect-ratio scaling is not a suitable technique for the
determination of $d_s$, at least for the system sizes accessible here: for the sizes
$L=3,\ldots,9$ of the $R=6$ systems, the wall simply has no space to properly
``meander'' over the length of this couple of lattice spacings, resulting in very
strong finite-size effects in $L$.  Finally, as the third combination of boundaries
considered, the O/DW spin boundary setup yields an extremely weak signal for the
chiral domain walls since improperly O($n$) rotated domains occur only scarcely
there, and we hence cannot produce a reliable estimate.  We take the average of the
P/AP and O/DW chiral results with sufficiently strong signals as our final estimate,
$d_s^\chi = 1.425(12)$. The analysis of spin domain walls from the overlap
$q_{l,\phi}^{AB}$ analogous to Eq.~(\ref{eq:rotation_overlap}) is more involved due
to the general dependence of the results on the cutoff angle $\phi_0$: if it is
chosen too small, smooth rotations are counted as parts of domain boundaries,
ultimately leading to space-filling domain walls in the limit $\phi_0 \rightarrow 0$.
Choosing it too large, on the other hand, misses to recognize some of the cluster
boundaries. We find somewhat stable results, however, in a range $0.5 \lesssim \phi_0
\lesssim 0.75$, and estimate $d_s^\phi =1.2379(97)$ for P/AP and $d_s^\phi =
1.251(21)$ for O/DW spin boundaries.  The O/DW chiral conditions yield too weak a
signal for the determination of $d_s^\phi$, and we again cite the average $d_s^\phi =
1.240(21)$ of the remaining two sets of boundaries as our best estimate. Thus, to the
extent that finite-size corrections and the cutoff dependence of $d_s^\phi$ are under
control, it appears that spin and chiral domain walls have different fractal
properties.


\section{Overlaps\label{sec:overlaps}}

One of the most striking results of the mean-field theory of the SK model is the
occurrence of a multitude of pure states, resulting in a non-trivial distribution of
the (thermally and disorder) averaged overlap parameter \cite{mezard:84}.
Consequently, estimates of the form of the overlap distribution have been regularly
considered as crucial benchmarks of the extent to which replica symmetry breaking
occurs in finite-dimensional spin glasses \cite{kawashima:03a}. In recent years it
has become clear, however, that great care has to be taken when drawing conclusions
from measurements on finite systems for the structure of pure states in spin glasses
\cite{middleton:99a,palassini:99,marinari:00a,newman:03a}. According to basic
principles of statistical mechanics \cite{ruelle:69}, a {\em thermodynamic state\/}
should be defined as a limiting probability measure from a series of systems of
increasing size, where a specific state is selected by an accompanying series of BCs.
The resulting state is then identified by the values in this limit (if they exist) of
all possible correlation functions in a central volume of the system, far away from
the perturbing presence of the boundaries. If such a state cannot be decomposed into
other thermodynamic states, i.e., it is mixed, it is termed pure. While this is
rather clear in homogeneous systems such as ferromagnets, unfortunately, it is not
known how a pure state should be explicitly constructed in spin-glass systems
\cite{newman:03a}.  In particular, there is ``chaotic size dependence'' in such a
sequence of system sizes resulting in oscillating and non-convergent correlation
functions. It is also unknown how a convergent subsequence could be selected by
coupling-independent BCs.  It is clear, then, that considering the overlap
distribution in whole, finite systems is not an appropriate way to detect the
presence of a multitude of pure states \footnote{Additionally, for the case of models
  without accidental degeneracies at $T=0$, it is clear that the optimized overlap
  $\hat{q}^{AB}$ is unity for each disorder configuration leading to a trivial
  overlap distribution $P^{AB}(q) = \delta(q-1)$ and hence does not reveal any
  information about the structure of the general state space.}. In particular, the
presence of domain walls separating patches of pure state configurations in finite
systems might lead to a non-trivial overlap distribution even if only a single class
of pure states related by a global O($n$) transformation was to exist. If, on the
other hand, spins in a volume far away from the boundaries always converge to the
same configuration (up to a common rotation) as $L\rightarrow\infty$, independent of
BCs, a multitude of nonequivalent and observable pure states would appear to be ruled
out.

An asymptotic independence from BCs of spin configurations in central windows can be
detected either in (window) overlap distributions or in the behavior of correlation
functions. These approaches are essentially equivalent \cite{palassini:99}. To see
this, consider two spins $S_i$ and $S_j$ inside a window $W$ containing $N_W = w^2$
spins (for simplicity we consider Ising spins first). The average squared difference
in the two-point functions between two systems with different BCs labelled $A$ and
$B$,
\begin{equation}
  \label{eq:twopoint_difference}
  \Delta = \frac{1}{N_W^2}\sum_{i,j\in W}\l[\l S_i^A S_j^A\r_T - \l S_i^B
  S_j^B\r_T]^2\r_J,
\end{equation}
by construction scales to zero at fixed window size $w$ if and only if the two-point
functions for all window spins agree for almost-every disorder configuration (here
$\l\cdot\r_T$ denotes a thermal and $\l\cdot\r_J$ a disorder average). It is then
easy to see that in terms of the optimized overlap $\hat{q}^{AB}$ of window spins
(which for the Ising case $n=1$ is just $|q^{AB}|$) this is equivalently written as
\begin{equation}
  \label{eq:window_overlap}
  \Delta = \l (\hat{q}^{AA})^2+(\hat{q}^{BB})^2-2(\hat{q}^{AB})^2\r_{T,J}.
\end{equation}
For the case of our {\em XY\/} model with a unique state at $T=0$, up to a global
O($n$) rotation (see Sec.~\ref{sec:spectrum} above), $\hat{q}^{AA}$ and
$\hat{q}^{BB}$ are of course always unity. Higher order (even) correlation functions
lead to similar expressions with higher moments of $\hat{q}^{AB}$, such that in the
most general context, the vanishing of the difference $\Delta P^{AB}(\hat{q}) =
P^{AA}(\hat{q})+P^{BB}(\hat{q})-2P^{AB}(\hat{q})$ of overlap distribution functions
is equivalent to the invariance of window correlation functions under a change of
BCs. For the present case of non-degenerate ground states, this vanishing is attained
only for the trivial form $P^{AB}(q)=\delta(\hat{q}-1)$, i.e., if and only if the
window configurations are completely unchanged for almost every disorder
configuration. A calculation along the same lines establishes the analogous relation
between correlation functions and overlaps for the case of continuous spins.

\begin{figure}[tb]
  \centering
  \includegraphics[clip=true,keepaspectratio=true,height=4.5cm]{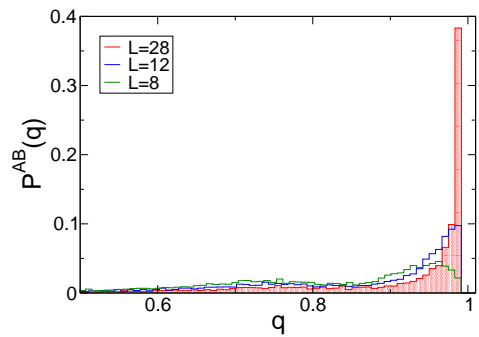}
  \caption
  {(Color online) Sampled distribution of optimized overlaps $\hat{q}^{AB}$ in
    central windows of size $w = 4$ for a change of BCs from periodic to
    antiperiodic.  As the system size is increased, the distribution accumulates all
    its weight close to $\hat{q}^{AB} = 1$. }
  \label{fig:overlap_histo}
\end{figure}

We did not attempt to sample ``all'' BCs, but used the P/AP, R/AR and O/DW boundaries
already considered for the defect-energy calculations. If the size $w$ of the overlap
window comes too close to the system size $L$, a crossover to the behavior of
whole-system overlaps is seen, such that one strives to guarantee $w \ll L$. For too
small windows, on the other hand, discretization or finite-size effects appear, and
there is a certain probability that the two window configurations can be made to
(almost) perfectly overlap by an O($n$) rotation although they are affected by the
influence of the changed boundaries --- down to the extreme case of $w=1$ where {\em
  each\/} pair of configurations has $\hat{q}^{AB} = 1$. Thus, the intended behavior
should appear for window sizes $1\ll w\ll L$. As we shall show below, with available
system sizes, these restrictions are best fulfilled for about $4\lesssim w\lesssim
8$, but finite-size effects are still clearly visible due to the maximum $L\approx
28$ that we can reach with the GEM. In Fig.~\ref{fig:overlap_histo} we present the
sampled distribution function of window overlaps $\hat{q}^{AB}$ for the P/AP
combination of BCs with $w=4$, showing a clear tendency of statistical weight to
accumulate at $\hat{q}^{AB} = 1$ as the system size $L$ is increased, suggesting a
convergence to the trivial form $P^{AB}(q)=\delta(\hat{q}-1)$ in the limit
$L\rightarrow\infty$. The overlap distributions for the remaining pairs of BCs show a
qualitatively similar behavior.  For a more systematic study of this apparent
convergence, we consider the total probability $p^{AB}(L,w)$ for the occurrence of an
overlap $\hat{q}^{AB}$ {\em smaller\/} than a threshold value $\hat{q}_0$, measuring
the distribution weight away from $\hat{q}^{AB} = 1$. To the extent to which the
configurational changes induced by a change of BCs take the form of domain walls,
this just measures the probability of the window being crossed by one of them. As
discussed in Sec.~\ref{sec:fractal}, such domain walls are fractal curves occupying a
volume $\sim L^{d_s}$, such that a fraction of the lattice proportional to
$L^{-(d-d_s)}$ will be touched by a domain wall (whether the relevant fractal
dimension would rather be connected with spin or chiral defects will have to be
decided post factum, see the discussion below). Since the relevant length scale in
fact is $L/w$ (remember that fractals are scale invariant), one expects an asymptotic
scaling \cite{middleton:99a}
\begin{equation}
  \label{eq:crossing_scaling}
  p^{AB}(L,w) = Q_0 (L/w)^{-(d-d_s)}.
\end{equation}
\begin{figure}[tb]
  \centering
  \includegraphics[clip=true,keepaspectratio=true,height=4.5cm]{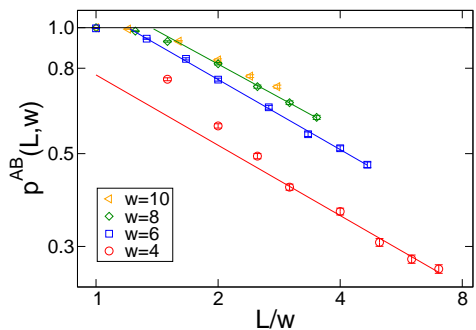}
  \caption
  {(Color online) Scaling of the probability $p^{AB}(L,w)$ of an overlap
    $\hat{q}^{AB} < 0.9$ for the P/AP pair of BCs as a function of $L/w$. The lines
    show fits of the form (\ref{eq:crossing_scaling}) at fixed $w$ to the data.  }
  \label{fig:overlap_scaling}
\end{figure}
Figure \ref{fig:overlap_scaling} shows the scaling of the crossing probability
$p^{AB}(L,w)$ for $\hat{q}_0 = 0.9$ and the P/AP pair of BCs. We additionally
considered cutoffs $\hat{q}_0 = 0.95$ and $\hat{q}_0 = 0.99$, changing the prefactor
$Q_0$ of (\ref{eq:crossing_scaling}) and slightly modifying corrections, but yielding
the same leading $(L/w)^{-(d-d_s)}$ behavior. There is a clear signal of scaling of
$p^{AB}(L,w)$ to zero as the system size $L$ is increased at fixed window size $w$.
The expected collapse of data for different window sizes, however, only appears at
$w\gtrsim 8$, at the edge of the window sizes that can be reasonably considered with
the system sizes at hand. From fits to the scaling form (\ref{eq:crossing_scaling})
we find exponent estimates $d_s = 1.442(29)$, $d_s = 1.444(16)$ and $d_s = 1.472(21)$
for window sizes $w = 4$, $w = 6$ and $w = 8$, respectively. These are not only
consistent with each other, but also compatible with the estimate $d_s^\chi =
1.425(12)$ found above for the fractal dimension of chiral domain walls. The
remaining combinations of BCs yield rather similar results, with somewhat more
pronounced finite-size corrections for the $L\le 20$ systems considered there (in
contrast to the aspect-ratio scaling of the domain-wall energies, where P/AP
boundaries entailed the stronger corrections). In particular, we find $d_s =
1.404(17)$ from R/AR boundaries ($w=6$), $d_s = 1.491(28)$ from O/DW spin BCs ($w=4$)
and $d_s = 1.489(59)$ for the O/DW chiral combination ($w = 6$).  It thus appears
that the relevant domain walls seen in the window overlaps are of the chiral type. In
total, clear signals of a scaling of the weight away from $\hat{q}^{AB} = 1$ to zero
are seen in all cases, indicating a ``deflection to infinity'' \cite{newman:03a} of
domain walls in the center of sufficiently large samples. The estimates for the
fractal dimension seem to preclude an excitation of space-filling domain walls (i.e.,
$d_s = 2$) as predicted by a many-state picture at least from the
coupling-independent and symmetric BCs considered. This is in agreement with the
findings for the 2D Ising spin glass of
Refs.~\onlinecite{palassini:99,middleton:99a}.


\section{Magnetic properties\label{sec:further}}
\subsection{Internal Fields}

\begin{figure}[tb]
  \centering
  \includegraphics[clip=true,keepaspectratio=true,height=5.5cm]{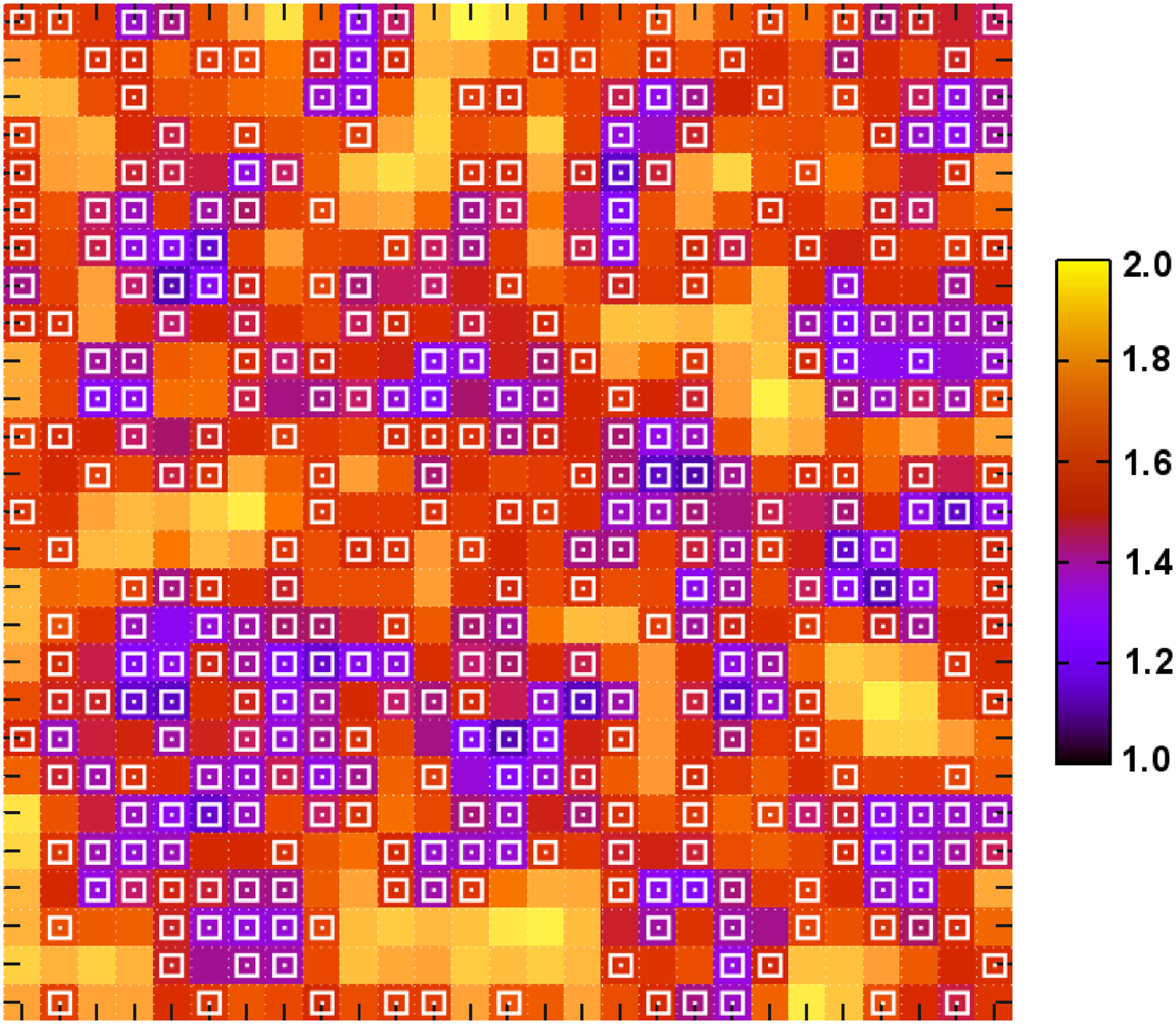}
  \caption
  {(Color online) Spatial distribution of the internal field per link in the ground
    state of an example $28\times 28$ disorder configuration with open-periodic BCs.
    There is a clear inverse correlation with the density of frustrated plaquettes
    (small white squares).}
  \label{fig:fields_density}
\end{figure}

It is straightforwardly seen that a necessary condition for the Hamiltonian
(\ref{eq:EA_model}) to be in a metastable state is that each spin is parallel to its
local internal field
\begin{equation}
  \label{eq:internal_field}
  \bm{h}^l_i = \frac{1}{q_i}\sum_j J_{ij}\bm{S}_j,
\end{equation}
which for convenience has been normalized to represent the field {\em per link\/} by
dividing by the local co-ordination number $q_i$, accounting for the difference in
connectivity between spins along open boundaries and in the bulk. An iterative
alignment of spin orientations along the internal field directions indeed drives the
system towards a metastable (but not a ground) state, and the corresponding
minimization technique is referred to as the Walker-Walstedt or spin-quench
algorithm \cite{walker:80}. The local strength of internal fields is closely linked
to the presence of frustration in the system, with a weak internal field indicating a
local uncertainty of the system with respect to choosing a given spin's orientation
such as to minimize the total energy. We indeed find a strong inverse correlation
between the size $|\bm{h}^l_i|$ of the internal fields in the ground state and the
density of frustrated plaquettes of the system. This is illustrated for the ground
state of an example disorder configuration of $28\times 28$ spins in
Fig.~\ref{fig:fields_density}. It is clear that in terms of finding a ground state,
optimal spin orientations in regions of low frustration (and consequently large
internal fields) are rather easily determined, while it is much harder to optimize
highly frustrated areas. This connection was exploited in the improved spin-quench
method suggested in Ref.~\onlinecite{gawiec:91a}. One then also expects a certain
correlation of the boundaries of rigid clusters discussed above in
Sec.~\ref{sec:gs_energy} and the occurrence of a high density of frustrated
plaquettes.

\begin{figure}[tb]
  \centering
  \includegraphics[clip=true,keepaspectratio=true,height=4.5cm]{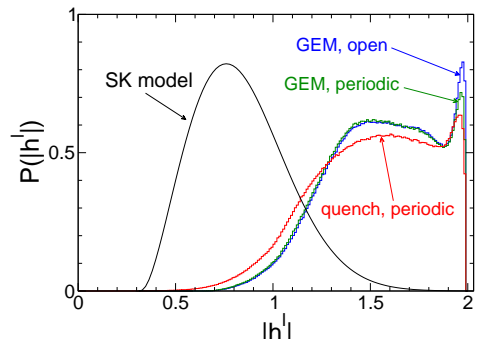}
  \caption
  {(Color online) Average distribution of internal fields $|\bm{h}^l|$. The ``GEM''
    curves represent ground states of $28\times 28$ systems with open-open and
    open-periodic BCs. The ``quench'' data correspond to the average over a broad
    range of metastable states as found from the Walker-Walstedt algorithm
    \cite{walker:80}.  The curve labeled ``SK model'' shows the corresponding
    distribution in the O($2$) SK model \cite{bray:81}.  }
  \label{fig:fields_dist}
\end{figure}

Speculations on the occurrence of a certain proportion of sites with almost or
exactly vanishing internal fields played some role in early models of the specific
heat in spin glasses, see Ref.~\onlinecite{walker:80} and references therein. For the
infinite-range case, it was subsequently established that $|\bm{h}^l_i| \ge h^l_0 >
0$, such that (almost) free spins do not occur \cite{bray:81}. For the short-range
Heisenberg model, a similar exact bound was reported in
Refs.~\onlinecite{kaplan:82,kaplan:82a}.  For the bimodal Ising spin glass, on the
other hand, a multitude of flippable spins exists in each ground state
\cite{hartmann:00}. For the intermediate {\em XY\/} case, however, to our knowledge
no exact results are available. Figure \ref{fig:fields_dist} shows the distribution
of internal fields $|\bm{h}^l|$ for $28\times 28$ systems. Independent of BCs, no
spins with fields $|\bm{h}^l| \lesssim 0.6 = h^l_0$ occur.  Smaller systems have
slightly larger gaps $h^l_0$, but the size dependence is very weak such that it is
clear that a finite $h^l_0$ survives in the thermodynamic limit. The absence of
almost free spins in the ground state clearly illustrates the fact that there are no
localized low-energy excitations and, instead, the lowest excitations are the
extended defects corresponding to domain walls or O($n$) rotations of rigid clusters,
as discussed in Sections~\ref{sec:gs_energy} and \ref{sec:walls}. The distributions
$P(|\bm{h}^l|)$ for the ground state show a characteristic behavior with the bulk of
spins having $1.25 \le |\bm{h}^l_i| \le 1.75$ indicating strong frustration, but a
certain number of almost fully satisfied spins leading to an extra peak close to
$|\bm{h}^l_i| \approx 2$ associated to regions free of frustrated plaquettes. There
are only small differences between the systems with open and periodic BCs, most
noticeably a relative enlargement of this peak of unfrustrated spins for the open
boundaries resulting from the relatively lower frustration of the extra boundary spins.
For comparison, we also show the corresponding distribution for states found from the
Walker-Walstedt zero-temperature quench, which asymptotically should result in an
even sampling of {\em all\/} metastable states. This leads to a rather different
distribution of internal fields where much less of the inherent frustration has been
relieved. These curves for the finite-range model are markedly different from the
distribution (again over all metastable states) of the infinite-range O($2$) SK model
\cite{bray:81} shown for comparison which, due to the lack of local fluctuations in
this infinite-dimensional model, is much more symmetric and more evenly frustrated.

\subsection{Magnetic phase diagram}

\begin{figure}[tb]
  \centering
  \includegraphics[clip=true,keepaspectratio=true,width=7cm]{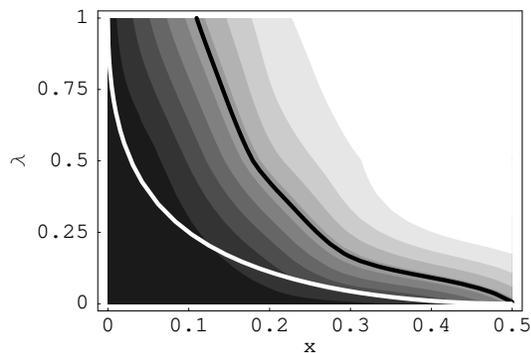}
  \caption
  {Contour plot of the average ground-state magnetization $\l|\bm{m}|\r_J$ for
    $16\times 16$ systems with open BCs as a function of $x$ and $\lambda$. Darker
    shades indicate larger magnetizations, and contours are separated by $\delta m =
    0.1$. The bold black line denotes the locus of the curve $\l|\bm{m}|\r_J = 0.5$,
    and the bold white line corresponds to the CPA prediction of Eq.~(\ref{eq:cpa}).}
  \label{fig:phase_diagram}
\end{figure}

For small concentrations $x$ and/or rather weak impurity bonds, one might expect the
system to exhibit some rather extended short-range order \cite{gawiec:91a}, or
perhaps even long-range order \cite{ching:90}, and from example calculations we
indeed find clearly non-zero magnetizations for this region, whereas strongly
disordered states of spin-glass signature appear once a larger number of strong
impurity bonds is being introduced. Analytical calculations regarding the question of
order and canting induced by impurity bonds have only been possible for isolated
defects or using the mean-field type coherent potential approximation (CPA) of
unclear precision for the case at hand \cite{saslow:88,parker:88,vannimenus:89}. As a
preliminary exploration of the zero-temperature phase diagram, we consider the
modulus of the magnetization, $\l|\bm{m}|\r_J$, where $\bm{m} = (\sum_i S_i)/N_0$.
Figure \ref{fig:phase_diagram} shows a contour plot of $\l|\bm{m}|\r_J$ as a function
of $x$ and $\lambda$ [cf.\ Eq.~(\ref{eq:coupling_distribution})] as found from GEM
ground-state computations for systems of $16\times 16$ spins and open BCs.  Numerical
calculations were actually performed at the grid of 30 points indicated by the tick
marks in the plot, and the contours were produced from a 3$^\mathrm{rd}$ order
polynomial interpolation of the results. The bold black line indicates the locus of
the curve $\l|\bm{m}|\r_J = 0.5$ which could be used as a possible definition of a
finite-size approximation of the phase transition line, approaching the true
transition line as $L\rightarrow\infty$. The prediction of the CPA, namely
\cite{vannimenus:89}
\begin{equation}
  \label{eq:cpa}
  x_c(\lambda) = \frac{1}{2} \frac{(1-\sqrt{\lambda})^2}{1+\lambda},
\end{equation}
is indicated by the bold white line. In the limit $\lambda \rightarrow 0$ of a
diluted and unfrustrated system, order must disappear at the (bond) percolation
threshold $x_c(\lambda = 0) = 0.5$ which is (by chance) already very well reproduced
by the chosen definition $\l|\bm{m}|\r_J = 0.5$ at the system size displayed. Also,
$x_c(0) = 1/2$ for the CPA form (\ref{eq:cpa}). There is no agreement in the
literature about the general form of $x_c(\lambda)$ as $x\rightarrow 0$, i.e., in the
experimentally most relevant regime of small dilution. According to (\ref{eq:cpa}),
the CPA predicts a finite intercept $\lambda_c = 1$ for $x = 0$, and this claim was
supported by the interpretation of the average magnetization found from a local
spin-quench computation in Ref.~\onlinecite{gawiec:91a} (at the same time, however,
further results from the same study seemed to indicate the absence of long-range
order for the whole phase diagram).  An RG calculation for the non-linear $\sigma$
model \cite{kruger:02}, on the other hand, appeared to indicate $\lambda_c
\rightarrow \infty$ as $x\rightarrow 0$. The results of Fig.~\ref{fig:phase_diagram}
certainly rather suggest that $\lambda_c > 1$ for $x\rightarrow 0$, but this
impression might be distorted by finite-size effects.  The question of long-range
order is here rather subtle \cite{vannimenus:89}, and it certainly cannot be excluded
from the data shown that the indicated transition line is related to a crossover
rather than a phase transition. In any case, this picture, of course, needs to change
again at finite temperatures, where true long-range order is impossible due to the
continuous symmetry of the system. A more thorough, finite-size scaling analysis of
this question, together with an investigation into the nature of the apparently
ordered phase (i.e., whether it is accompanied by transverse spin-glass order
\cite{gawiec:91a}), is left as an interesting future application of the GEM
technique.

\section{Conclusions\label{sec:concl}}

We have shown that a newly developed specially tailored heuristic optimization
technique combining exact ground state computations for embedded Ising variables with
a genetic algorithm allows for a thorough investigation of the ground state of the
planar EA spin glass in two dimensions, contributing to a resolution of a number of
long-standing issues for this system.

The main result consists of a careful determination of the spin and chiral stiffness
exponents of the model with bimodal coupling distribution, which characterize its
critical behavior and directly relate to the question of a possible spin-chirality
decoupling in this system. In principle, a stiffness exponent related to domain walls
can be extracted from scaling corrections to the finite-size behavior of the
ground-state energy, but the precision of the estimate resulting from this method is
not satisfactory.  Instead, we investigated the properties of domain-wall defects
directly induced by a suitable change of boundary conditions. In response to doubts
regarding the suitability of the conventionally considered combination of periodic
and antiperiodic boundary conditions resulting in an effective measurement of {\em
  differences\/} of defect energies \cite{kosterlitz:99a}, we explicitly excite
domain-wall defects of the spin and chiral type relative to a system with free
boundaries by applying special ``domain-wall boundary conditions''. Corrections to
scaling for these as well as for the additionally considered periodic-antiperiodic
and random-antirandom boundaries are found to be large and their resolution, from
explicitly including correction terms in fits to the defect energies, is not fully
satisfactory given the accessible system sizes. However, application of the
aspect-ratio scaling technique \cite{carter:02a} allows for a good control over the
boundary-dependent corrections, leading to a consistent estimate $\theta_s =
-0.329(14)$ of the spin stiffness exponent for the different sets of boundary
conditions used. This value is considerably smaller in modulus than the results of
previous authors, quoting values of $\theta_s = -1.0$,\cite{jain:86a} $-0.9$,
\cite{morris:86a} $-0.8$, \cite{kawamura:91} $-0.78$, \cite{maucourt:98a} and
$-0.37(2)$, \cite{kosterlitz:99a} respectively. The spread of these previous results
nicely illustrates the magnitude of present finite-size corrections as well as the
hardness of computing true ground states for sufficiently large systems. The fact
that in the present work, consistency between estimates of the stiffness exponent
from different sets of boundary conditions with correction amplitudes of opposite
signs could be achieved, makes it appear plausible that finally scaling corrections
are under control to an acceptable degree of precision. Concerning the claim put
forward in Ref.\ \onlinecite{kosterlitz:99a} that the P/AP combination of boundary
conditions cannot be used to estimate the spin stiffness exponent, the results of the
aspect-ratio scaling show that the assumed trapping of domain walls in periodic
systems \cite{kosterlitz:99a} only affects scaling corrections, but does not alter
the asymptotic stiffness. This value of $\theta_s$ is remarkably close, but
apparently not identical to the estimate $\theta_s = -0.287(4)$ found for the
two-dimensional Ising spin glass with {\em Gaussian\/} coupling distribution
\cite{hartmann:02a}.  Also, scaling corrections to the ground state energy are found
to be very close to the Gaussian Ising model, but rather different from the bimodal
Ising spin glass.  Chiral boundaries, on the other hand, lead to a clearly different
exponent $\theta_c = -0.114(16)$, which is again remarkably smaller in modulus than
previous estimates of $\theta_c = -0.6$, \cite{bokil:96a} $-0.38$,
\cite{kawamura:91,maucourt:98a} and $-0.37(1)$, \cite{kosterlitz:99a} respectively.
Consequently, it appears that spin and chiral degrees-of-freedom order with different
exponents in this system although, of course, a study for finite system sizes can
never guarantee to probe the true asymptotic behavior. Naively, one might expect that
the Ising-like chiral variables should lead to the value $\theta_c = 0$ of the
bimodal Ising spin glass \cite{hartmann:01a}.  It should be noted, however, that the
chiral variables interact with an effective long-range Coulombic interaction
\cite{villain:77a}, possibly leading to a shift in universality class. On the other
hand, for the recently established spin-chirality decoupling in the periodically
frustrated {\em XY\/} model \cite{ozeki:03,hasenbusch:05}, the chiral transition
appears to show Onsager exponents, but this behavior is only seen for huge lattice
sizes of $L\approx 10^3$, with wildly different behavior in the crossover region
\cite{hasenbusch:05}.  For our data, since for a $\theta_c$ very close to zero it
becomes very hard numerically to distinguish the leading scaling from sub-leading
corrections, we cannot definitely exclude the possibility of $\theta_c = 0$. In fact,
including a finite asymptotic value $\Delta E_0$ in the form
(\ref{eq:dw_corrections}) fits consistent with $\theta_c = 0$ can be achieved as
well. In the context of the observed spin-chirality decoupling, it is therefore quite
interesting that we find a $\theta_s$ value so close to that of the 2D Ising spin
glass with Gaussian interactions.

We find that the structure of metastable states of the model can be understood in
terms of metastates of clusters of rigidly locked spins, which undergo common
relative O($n$) rotations between different metastable states. To each metastable
state belongs a spectrum of continuous spin-wave excitations leaving the associated
energy minimum. These will be considered in a future investigation. Low-energy
excitations {\em within\/} the manifold of metastable states, on the other hand, are
extended objects, corresponding to O($n$) rotations of one or several rigid clusters.
Almost free spins corresponding to local excitations do not occur. Many such
metastable configurations are found closely above the ground state, leading to a
quasi-continuous spectrum in the thermodynamic limit. The ground states themselves,
however, are found to be unique up to a global O($n$) transformation, implying a
critical exponent $\eta = 0$ of the ground-state correlation function. Together with
the estimates for $\theta_s$ and $\theta_c$, this gives a rather complete
characterization of the $T=0$ critical point of the model.

We investigated the nature of elementary excitations induced by a change of boundary
conditions by analyzing the locally averaged optimal rotation matrices between
ground-state configurations. Chiral domain walls are by definition sharply localized.
We find them to be fractal curves, and from the scaling of the corresponding link
overlaps we estimate a fractal dimension $d_s^\chi = 1.425(12)$. The definition of
spin domain walls depends on a somewhat arbitrary cutoff angle, but we find the
fractal dimension $d_s^\phi = 1.240(21)$ to be independent of the cutoff within some
reasonable range.  Recently, evidence was presented for the hypothesis that domain
walls in the Ising spin glass are special fractal curves described by ``stochastic
Loewner evolution'' (SLE) \cite{amoruso:06a,bernard:06}. Comparing the numerical
results with conformal weights of the Kac table, the authors of
Ref.~\onlinecite{amoruso:06a} conjecture a relation $d_s = 1+3/4(3+\theta)$. This we
find fulfilled within statistical errors for $d_s^\phi$, but not for the chiral
exponent $d_s^\chi$ where one might have rather expected it to hold. In any case,
from the excitation via a change of boundary conditions, we do {\em not\/} find
space-filling domain walls ($d_s = 2$) one necessarily expects for a model with many
thermodynamic pure states \cite{newman:03a}. This appears to be corroborated by an
explicit investigation of the distribution of overlaps in central windows far away
from the boundaries, which, for the system sizes accessible here, appears to converge
to a trivial form $P^{AB}(\hat{q}) = \delta(\hat{q}-1)$. This shows that the states
defined by the considered set of boundary conditions always converge to the same
class of thermodynamic pure states related by a global O($n$) rotation, and we do not
find evidence for a multitude of pure states. Naturally, however, consideration of
different types of boundary conditions might yield different results. The rate of
this convergence is again related to a domain-wall fractal dimension, with a value
compatible with $d_s^\chi$ as determined directly.

A multitude of extensions of the present work comes to mind. Besides the mentioned
investigation of spin-wave excitations complementing the information on metastable
states presented here, certainly the magnetic phase diagram discussed in
Sec.~\ref{sec:further}, which is relevant to a number of experimentally realized
systems, deserves a more extensive investigation. Concerning the structure of the
ground and excited states in terms of clusters of rigid spins, a more systematic
analysis of the morphology along the lines of Refs.~\onlinecite{barahona:82a,roma:06}
appears promising.  Clearly, investigations of Gaussian bond distributions and of the
Heisenberg case would be highly interesting enterprises. Above all, however, the
important question of spin-chirality decoupling in three-dimensional spin glasses
could be tackled with a suitable generalization of the GEM approach employed here,
replacing the embedded matching component inside the genetic algorithm by a technique
applicable to three-dimensional systems.



\begin{acknowledgments}
  The authors acknowledge fruitful discussions with I.\ Campbell. The research at the
  University of Waterloo was undertaken, in part, thanks to funding from the Canada
  Research Chairs Program (M.G., Tier 1), the Canada Foundation for Innovation and
  the Ontario Innovation Trust (M.G.). M.W.\ acknowledges support by the EC in form
  of a ``Marie Curie Intra-European Fellowship'' under contract No.\
  MEIF-CT-2004-501422. This work was made possible, in part, by the facilities of the
  Shared Hierarchical Academic Research Computing Network (SHARCNET:
  www.sharcnet.ca).
\end{acknowledgments}



\end{document}